\RequirePackage{fix-cm}

\documentclass[twocolumn,epjc3]{svjour3}
\smartqed % flush right qed marks, e.g. at end of proof
\usepackage{textcomp}% for '\textdegree' macro
\RequirePackage{graphicx}
\RequirePackage[numbers,sort&compress]{natbib}
\journalname{Eur.\ Phys.\ J.\ C}
\usepackage{dcolumn} % Include tables packages
\usepackage{latexsym}
\usepackage{graphics}
\usepackage{amsmath}
\usepackage{wasysym}
\usepackage{subfig}
\usepackage{color}
\usepackage{mhchem}
\usepackage{siunitx}
\usepackage[svgnames]{xcolor}

\newcommand{\po}{\ce{^{210}Po}}
\newcommand{\bi}{\ce{^{210}Bi}}
\newcommand{\pb}{\ce{^{210}Pb}}
\newcommand{\be}{\ce{^{7}Be}}

\newcommand{\celeven}{\ce{^{11}C}}
\newcommand{\pep}{{\it pep}}
\newcommand{\pp}{{\it pp}}
\newcommand{\cpd}{cpd$/100$ ton}
\newcommand{\red}[1]{{\color{black} #1}}

\newcommand{\mwe}{m.w.e.}

\usepackage{caption}
\usepackage{amssymb}
\usepackage{graphicx}
\usepackage{sidecap}
\usepackage{subfig}

\newcommand{\Borexino}{\textsc{BOREXINO collaboration\thanksref{spokes}}}
\newcommand{\APC}{AstroParticule et Cosmologie, Universit\'e Paris Diderot, CNRS/IN2P3, CEA/IRFU, Observatoire de Paris, Sorbonne Paris Cit\'e, 75205 Paris Cedex 13, France}
\newcommand{\Dubna}{Joint Institute for Nuclear Research, 141980 Dubna, Russia}
\newcommand{\Genova}{Dipartimento di Fisica, Universit\`a degli Studi and INFN, 16146 Genova, Italy}

\newcommand{\Krakow}{M.~Smoluchowski Institute of Physics, Jagiellonian University, 30348 Krakow, Poland}
\newcommand{\Kiev}{Kiev Institute for Nuclear Research, 03680 Kiev, Ukraine}
\newcommand{\Kurchatov}{National Research Centre Kurchatov Institute, 123182 Moscow, Russia}
\newcommand{\Kurchatovb}{National Research Nuclear University MEPhI (Moscow Engineering Physics Institute), 115409 Moscow, Russia}
\newcommand{\LNGS}{INFN Laboratori Nazionali del Gran Sasso, 67010 Assergi (AQ), Italy}
\newcommand{\Milano}{Dipartimento di Fisica, Universit\`a degli Studi and INFN, 20133 Milano, Italy}
\newcommand{\Perugia}{Dipartimento di Chimica, Biologia e Biotecnologie, Universit\`a degli Studi e INFN, 06123 Perugia, Italy}
\newcommand{\Peters}{St. Petersburg Nuclear Physics Institute NRC Kurchatov Institute, 188350 Gatchina, Russia}
\newcommand{\Princeton}{Physics Department, Princeton University, Princeton, NJ 08544, USA}
\newcommand{\PrincetonChemEng}{Chemical Engineering Department, Princeton University, Princeton, NJ 08544, USA}
\newcommand{\UMass}{Amherst Center for Fundamental Interactions and Physics Department, University of Massachusetts, Amherst, MA 01003, USA}
\newcommand{\Virginia}{Physics Department, Virginia Polytechnic Institute and State University, Blacksburg, VA 24061, USA}
\newcommand{\Munchen}{Physik-Department and Excellence Cluster Universe, Technische Universit\"at M\"unchen, 85748 Garching, Germany}
\newcommand{\Lomonosov}{Lomonosov Moscow State University Skobeltsyn Institute of Nuclear Physics, 119234 Moscow, Russia}
\newcommand{\GSSI}{Gran Sasso Science Institute, 67100 L'Aquila, Italy}
\newcommand{\Dresda}{Department of Physics, Technische Universit\"at Dresden, 01062 Dresden, Germany}
\newcommand{\Mainz}{Institute of Physics and Excellence Cluster PRISMA$^+$, Johannes Gutenberg-Universit\"at Mainz, 55099 Mainz, Germany}

\newcommand{\Juelich}{Institut f\"ur Kernphysik, Forschungszentrum J\"ulich, 52425 J\"ulich, Germany}
\newcommand{\RWTH}{III. Physikalisches Institut B, RWTH Aachen University, 52062 Aachen, Germany}
\newcommand{\Aquila}{Dipartimento di Scienze Fisiche e Chimiche, Universit\`a dell'Aquila, 67100 L'Aquila, Italy}

\newcommand{\Napoli}{Present address: Dipartimento di Fisica, Universit\`a degli Studi Federico II e INFN, 80126 Napoli, Italy}

\newcommand{\LNGSG}{Present address: INFN Laboratori Nazionali del Gran Sasso, 67010 Assergi (AQ), Italy}
\newcommand{\California}{Present address: University of California, Berkeley, Department of Physics, CA 94720, Berkeley, USA}
\newcommand{\UniLondon}{Department of Physics, Royal Holloway, University of London, Department of Physics, School of Engineering, Physical and Mathematical Sciences, Egham, Surrey, TW20 OEX}
\newcommand{\Madrid}{Present address: Departamento de F\'{i}sica Te\'{o}rica, Universidad Aut\'{o}noma de Madrid, Campus Universitario de Cantoblanco, 28049 Madrid, Spain}
\newcommand{\spokes}{Corresponding author: spokesperson-borex@lngs.infn.it}

\begin{document}
\title{Sensitivity to neutrinos from the solar CNO cycle in Borexino}
\author{
M.~Agostini\thanksref{Munchen}\and
K.~Altenm\"{u}ller\thanksref{Munchen}\and
S.~Appel\thanksref{Munchen}\and
V.~Atroshchenko\thanksref{Kurchatov}\and
Z.~Bagdasarian\thanksref{Juelich,California}\and
D.~Basilico\thanksref{Milano}\and
G.~Bellini\thanksref{Milano}\and
J.~Benziger\thanksref{PrincetonChemEng}\and
R.~Biondi\thanksref{LNGS}\and
D.~Bravo\thanksref{Milano,Madrid}\and
B.~Caccianiga\thanksref{Milano}\and
F.~Calaprice\thanksref{Princeton}\and
A.~Caminata\thanksref{Genova}\and
P.~Cavalcante\thanksref{Virginia,LNGSG}\and
A.~Chepurnov\thanksref{Lomonosov}\and
D.~D'Angelo\thanksref{Milano}\and
S.~Davini\thanksref{Genova}\and
A.~Derbin\thanksref{Peters}\and
A.~Di Giacinto\thanksref{LNGS}\and
V.~Di Marcello\thanksref{LNGS}\and
X.F.~Ding\thanksref{Princeton}\and
A.~Di Ludovico\thanksref{Princeton}\and
L.~Di Noto\thanksref{Genova}\and
I.~Drachnev\thanksref{Peters}\and
A.~Formozov\thanksref{Dubna,Milano}\and
D.~Franco\thanksref{APC}\and
C.~Galbiati\thanksref{Princeton,GSSI}\and
C.~Ghiano\thanksref{LNGS}\and
M.~Giammarchi\thanksref{Milano}\and
A.~Goretti\thanksref{Princeton,LNGSG}\and
A.S. G\"{o}ttel\thanksref{Juelich,RWTH}\and
M.~Gromov\thanksref{Lomonosov,Dubna}\and
D.~Guffanti\thanksref{Mainz}\and
Aldo~Ianni\thanksref{LNGS}\and
Andrea~Ianni\thanksref{Princeton}\and
A.~Jany\thanksref{Krakow}\and
D.~Jeschke\thanksref{Munchen}\and
V.~Kobychev\thanksref{Kiev}\and
G.~Korga\thanksref{UniLondon}\and
S.~Kumaran\thanksref{Juelich,RWTH}\and
M.~Laubenstein\thanksref{LNGS}\and
E.~Litvinovich\thanksref{Kurchatov,Kurchatovb}\and
P.~Lombardi\thanksref{Milano}\and
I.~Lomskaya\thanksref{Peters}\and
L.~Ludhova\thanksref{Juelich,RWTH}\and
G.~Lukyanchenko\thanksref{Kurchatov}\and
L.~Lukyanchenko\thanksref{Kurchatov}\and
I.~Machulin\thanksref{Kurchatov,Kurchatovb}\and
J.~Martyn\thanksref{Mainz}\and
E.~Meroni\thanksref{Milano}\and
M.~Meyer\thanksref{Dresda}\and
L.~Miramonti\thanksref{Milano}\and
M.~Misiaszek\thanksref{Krakow}\and
V.~Muratova\thanksref{Peters}\and
B.~Neumair\thanksref{Munchen}\and
M.~Nieslony\thanksref{Mainz}\and
R.~Nugmanov\thanksref{Kurchatov,Kurchatovb}
L.~Oberauer\thanksref{Munchen}\and
V.~Orekhov\thanksref{Kurchatov}\and
F.~Ortica\thanksref{Perugia}\and
M.~Pallavicini\thanksref{Genova}\and
L.~Papp\thanksref{Munchen}\and
\"O.~Penek\thanksref{Juelich,RWTH}\and
L.~Pietrofaccia\thanksref{Princeton}\and
N.~Pilipenko\thanksref{Peters}\and
A.~Pocar\thanksref{UMass}\and
G.~Raikov\thanksref{Kurchatov}\and
M.T.~Ranalli\thanksref{LNGS}\and
G.~Ranucci\thanksref{Milano}\and
A.~Razeto\thanksref{LNGS}\and
A.~Re\thanksref{Milano}\and
M.~Redchuk\thanksref{Juelich,RWTH}\and
A.~Romani\thanksref{Perugia}\and
N.~Rossi\thanksref{LNGS}\and
S.~Sch\"onert\thanksref{Munchen}\and
D.~Semenov\thanksref{Peters}\and
G. Settanta\thanksref{Juelich}\and
M.~Skorokhvatov\thanksref{Kurchatov,Kurchatovb}\and
O.~Smirnov\thanksref{Dubna}\and
A.~Sotnikov\thanksref{Dubna}\and
Y.~Suvorov\thanksref{LNGS,Kurchatov,Napoli}\and
R.~Tartaglia\thanksref{LNGS}\and
G.~Testera\thanksref{Genova}\and
J.~Thurn\thanksref{Dresda}\and
E.~Unzhakov\thanksref{Peters}\and
F.L.~Villante\thanksref{LNGS,Aquila}\and
A.~Vishneva\thanksref{Dubna}\and
R.B.~Vogelaar\thanksref{Virginia}\and
F.~von~Feilitzsch\thanksref{Munchen}\and
M.~Wojcik\thanksref{Krakow}\and
M.~Wurm\thanksref{Mainz}\and
S.~Zavatarelli\thanksref{Genova}\and
K.~Zuber\thanksref{Dresda}\and
G.~Zuzel\thanksref{Krakow} (\Borexino\label{Borexino})
}

\institute{\Munchen\label{Munchen}\and
\Kurchatov\label{Kurchatov}\and
\Juelich\label{Juelich}\and
\Milano\label{Milano}\and
\PrincetonChemEng\label{PrincetonChemEng}\and
\LNGS\label{LNGS}\and
\Princeton\label{Princeton}\and
\Genova\label{Genova}\and
\Virginia\label{Virginia}\and
\Lomonosov\label{Lomonosov}\and
\Peters\label{Peters}\and
\Dubna\label{Dubna}\and
\APC\label{APC}\and
\GSSI\label{GSSI}\and
\RWTH\label{RWTH}\and
\Mainz\label{Mainz}\and
\Krakow\label{Krakow}\and
\Kiev\label{Kiev}\and
\UniLondon\label{UniLondon}\and
\Kurchatovb\label{Kurchatovb}\and
\Dresda\label{Dresda}\and
\Perugia\label{Perugia}\and
\UMass\label{UMass}\and
\Aquila\label{Aquila}
}
\thankstext{spokes}{\spokes}
\thankstext{California}{\California}
\thankstext{Madrid}{\Madrid}
\thankstext{LNGSG}{\LNGSG}
\thankstext{Napoli}{\Napoli}

\date{Received: \today / Revised version: \today}
% The correct dates will be entered by Springer
%
%
\maketitle

%%%%%%%%%%%% Abstract %%%%%%%%%%%%
\abstract{Neutrinos emitted in the carbon, nitrogen, oxygen (CNO) fusion cycle in the Sun are a sub-dominant, yet crucial component of solar neutrinos whose flux has not been measured yet. The Borexino experiment at the Laboratori Nazionali del Gran Sasso (Italy) has a unique opportunity to detect them directly thanks to the detector's radiopurity and the precise understanding of the detector backgrounds. We discuss the sensitivity of Borexino to CNO neutrinos, which is based on the strategies we adopted to constrain the rates of the two most relevant background sources, \(pep\) neutrinos from the solar $pp$-chain and \(^{210}\)Bi beta decays originating in the intrinsic contamination of the liquid scintillator with \(^{210}\)Pb.
\\
 \red{Assuming the CNO flux predicted by the high-metallicity Standard Solar Model and an exposure of 1000\,days $\times$ 71.3\,t, Borexino has a median sensitivity to CNO neutrino higher than 3 \(\sigma\). With the same hypothesis the expected experimental uncertainty on the CNO neutrino flux is 23\%, provided the uncertainty on the independent estimate of the \bi{} interaction rate is 1.5 \cpd{}.}

 Finally, \red{we evaluated the expected uncertainty of the C and N abundances and the expected discrimination significance between the high and low metallicity Standard Solar Models (HZ and LZ) with future more precise measurement of the CNO solar neutrino flux.}
}

%%%%%%%%%%%% Introduction %%%%%%%%%%%%
\section{Introduction}
\label{sec:Intro}
\begin{figure}[t]
    \centering
    \includegraphics[width=0.9\columnwidth]{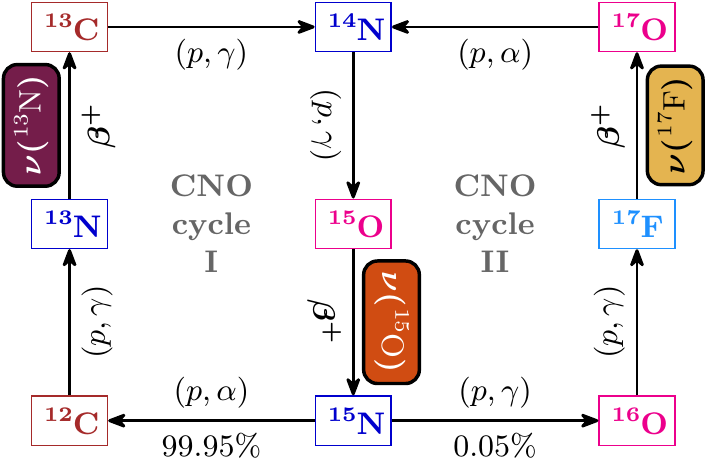}
    \caption{Two branches (CNO-I and CNO-II) of the CNO cycle of proton-proton fusion to \ce{^4He}. Only the former is complete in the Sun's core \cite{bahcall_book}.}
    \label{fig:CNOcycle}
\end{figure}

\begin{figure}[h]
    \centering
    \includegraphics[width=\columnwidth]{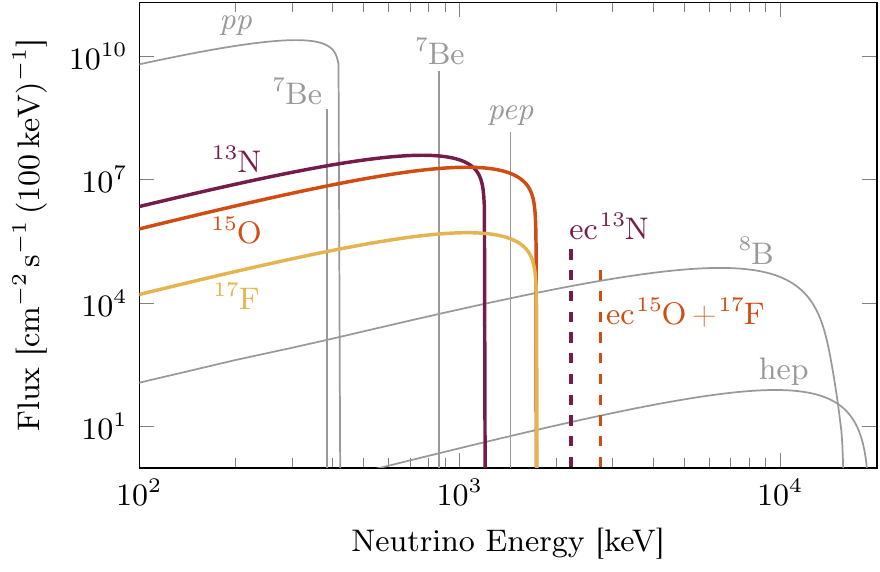}
    \caption{Solar neutrino spectra predicted by the \textit{B16 (GS98)-HZ} Standard Solar Model \cite{B16SSM}. For monochromatic lines, the vertical axis reports the flux in $\rm{cm^{-2} s^{-1}}$. The so-called \ce{^{13}N} and \ce{^{15}O} neutrinos are produced by the $\beta^+$-decays of Nitrogen-13 and Oxygen-15 in the CNO-I-branch of the CNO-cycle, respectively. The Fluorine-17 component is essentially negligible in the Sun.}
    \label{fig:SolarNuSpectrum}
\end{figure}

The Sun releases energy mainly through a nuclear fusion process known as the proton-proton chain (\pp{} chain). Another process, called the carbon-nitrogen-oxygen (CNO) cycle (see Fig. \ref{fig:CNOcycle}), is expected to contribute about one percent of the total energy and neutrino production \cite{clayton_book,bahcall_book}. The CNO cycle emits neutrinos with energies up to around \SI{1.2}{MeV} and \SI{1.7}{MeV} for its two main components, see Fig. \ref{fig:SolarNuSpectrum}. This process, thought to be the dominant energy production process for stars heavier than 1.3 solar masses \cite{Salaris} as well as solar-like stars in advanced evolutionary stages \cite{Kippenhan_Book}, has many implications for astrophysical problems. For example, the measurement of CNO neutrinos would allow the evaluation of the efficiency of the CNO cycle, helping with the determination of the age of globular clusters \cite{Degli_Innocenti}. It would also provide a direct reading of the metal abundance in the solar core, which would in turn allow the study of the chemical evolution paradigm assumed by the Standard Solar Model (SSM) \cite{B16SSM}.

Currently the CNO neutrino rate is only measured to be less than 8.1 \red{counts per day per 100\,ton (hereinafter as \cpd{})} at 95\% confidence level by Borexino \cite{ppchainNature}. This paper presents a detailed study of the sensitivity of the Borexino experiment to CNO neutrinos. The study relies on a precise and independent determination of the two main residual background components: $^{210}$Bi contamination {\color{black} of the liquid scintillator}, which can be estimated using measurements of $^{210}$Po decays (as suggested in \cite{PoBipaper}), and \pep{}  {\color{black} solar} neutrinos, which are a part of the \pp{} chain.

After a brief overview of the Borexino detector in Sec. \ref{sec:detector}, Sec. \ref{sec:CNOStrat:Sens} discusses Borexino's sensitivity to CNO neutrinos. \red{Sections \ref{sec:CNOStrat:Bkg} present the strategy to constrain the backgrounds and the influence of additional purification of the liquid scintillator.}  In Section~\ref{sec:SSMimpact} we comment on the relevance of the measurement of the flux of CNO neutrinos in the context of solar physics, with an emphasis on the ``solar metallicity (or abundance) problem''. This scientific puzzle originated when a re-determination of the surface metallicity of the Sun \cite{AGS05,AGSS09,AGSS15a,AGSS15b,Cob11} indicated a lower value than previously assumed~\cite{GS98}. However, solar models incorporating these lower metal abundances \cite{AGSS09} meet difficulties in reproducing the results from helioseismology, which support models with a higher metal content. Section \ref{sec:SSMimpact:CNAbundance} shows that, by following the approach proposed in \cite{Serenelli:2013,Haxton:2008}, it is possible to infer the carbon and nitrogen contents of the solar core independently of the assumed opacity of solar plasma by combining a CNO neutrino flux measurement with the very precise measurement of the \ce{^{8}B} neutrino flux by the Super-Kamiokande collaboration \cite{Abe2016} (about 2\% precision). Finally, Sec. \ref{sec:SSMimpact:HpTest} discusses the possibility of using a CNO neutrino measurement to discriminate among SSMs with different hypotheses about the Sun's surface metallicity by combining a measurement of the CNO neutrino flux with existing $^7$Be and $^8$B data measured by Borexino \cite{ppchainNature}.

\begin{figure*}[ht]
 \centering
 \includegraphics[width=0.8\textwidth]{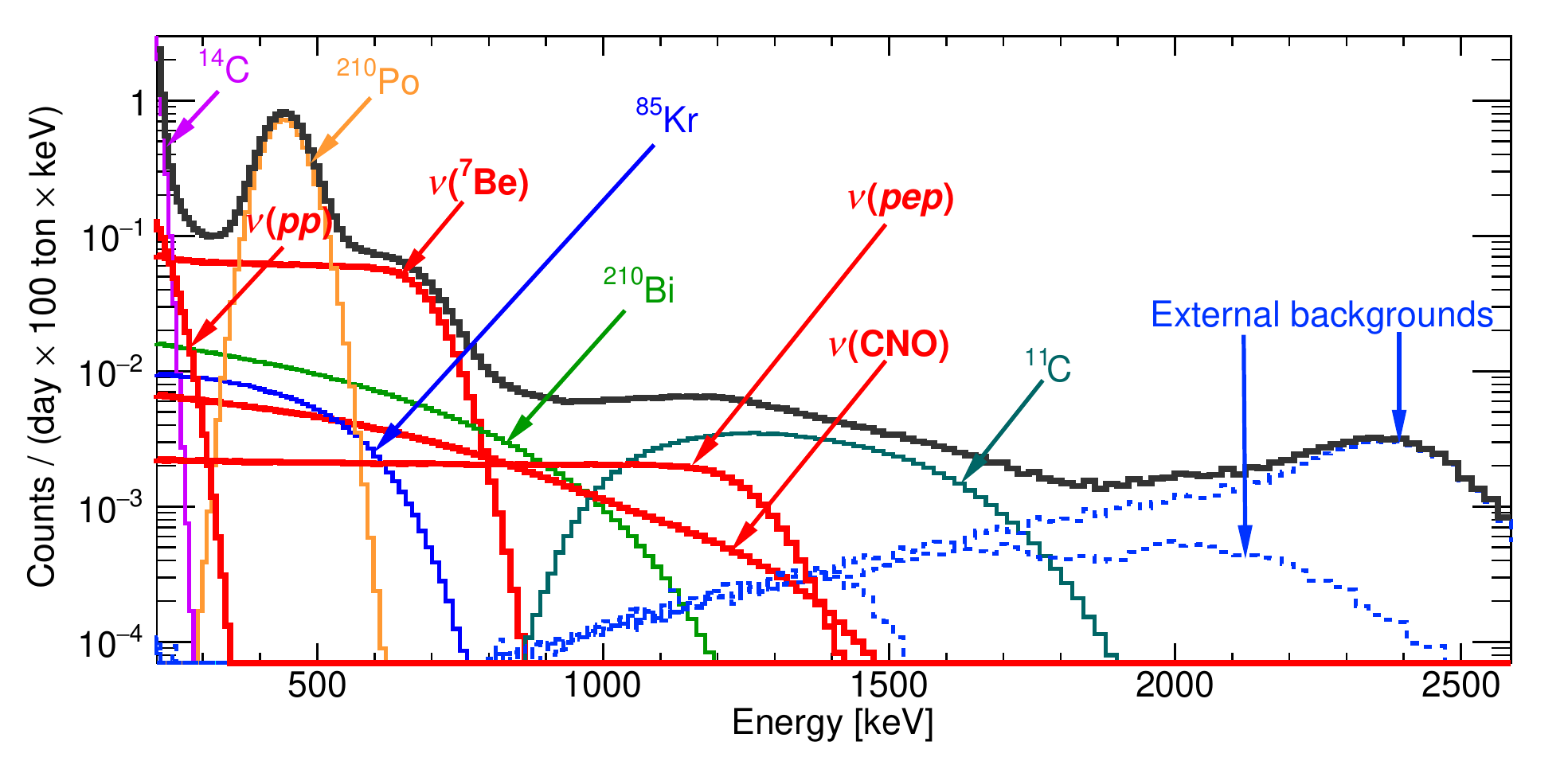}
 \caption
 {Expected event energy distribution of an arbitrary exposure without statistical fluctuations and after the fiducial volume cut.
 The solar neutrino components are highlighted with bold red lines. The spectra of the relevant background contributions are scaled to existing measurements. The contribution of the decay of the \ce{^{11}C} isotope reported in the plot is reduced by the application of the Three Fold Coincidence (TFC) procedure
 described in \cite{BxPhaseI,BxNuSol}.
 }
 \label{fig:BxSpectrum}
\end{figure*}

%%%%%%%%%%%% Experiment description %%%%%%%%%%%%
\section{The Borexino Experiment}
\label{sec:detector}
The Borexino detector \cite{Alimonti2009a} has been taking data since 2007 in the Laboratori Nazionali del Gran Sasso (Italy) at a depth of \SI{3800}{\mwe}
Borexino is an unsegmented calorimeter using about  {\color{black}280}\,ton of
ultra-pure organic liquid scintillator. The scintillator consists of PC (pseudocumene, 1,2,4-trimethyl-benzene) as a solvent with 1.5 g/l of fluor PPO (2,5-diphenyloxazole) as a solute. The electron density is \((3.307\pm0.015)\times 10^{31}\) e$^-$ per 100\,ton.
This scintillator is contained in a \SI{125}{\micro\meter}-thick nylon vessel with a nominal radius of \SI{4.25}{\meter}. It is surrounded by non-scintillating buffer fluid contained in a \SI{6.85}{\meter} radius stainless steel sphere, which supports 2212 8-inch ETL 9351 photomultipliers (PMTs). The stainless steel sphere is submerged in ultra-pure water serving as an active muon veto and a passive shield against external radiation.

Neutrinos are detected via the elastic scattering reaction on electrons in the liquid scintillator. The electrons deposit their energy in the scintillator, which results in scintillation light.%
 \red{To measure this deposited energy, several energy estimators are developed~\cite{BxPhaseI, BxNuSol} and they give consistent results in solar neutrino analyses. In this work, we use the variable \(N_h\), the overall number of hits on the PMTs, as a baseline.}
The interaction vertex is reconstructed via the arrival times of the scintillation photons on the PMTs. As a reference, the detected photoelectron yield is $\sim$500\,photoelectrons/MeV (normalized to 2000 PMTs). The energy and position reconstruction uncertainties at 1\,MeV are $\sim$\,50\,keV and $\sim$\,10\,cm, respectively.

The time distribution of the detected hits on the PMTs allows pulse-shape discrimination (PSD). With PSD, \(\alpha\) particles, a product of \po{} decay in the scintillator, can be discriminated from \(\beta\) particles on an event-by-event basis \cite{SeasonalModulation,Agostini2020x}. It can also be used to constrain the ratio of the total counts of \(\beta^+\) events, such as \ce{^{11}C} decays, to \(\beta^-\) events in specific energy ranges \cite{BxPhaseI, BxNuSol}.

%=====================================
\begin{table*}
\centering
\caption{Expected integral interaction rates (without energy threshold), and corresponding Borexino Phase-II results \cite{ppchainNature}, in \cpd{}.
 The prediction assumes 
 the Standard Solar Model \cite{B16SSM} under the high (HZ) and low (LZ) metallicity hypotheses
 and the MSW-LMA paradigm with the oscillation parameters reported in \cite{Esteban2017}.
 The data have been obtained with a total exposure of 1291.51 day$\times$71.3 ton.
 The CNO interaction rate was constrained to the expected HZ or LZ values in the fit procedure thereby allowing the measurement of the \pp{}, $^7$Be, and \pep{} interaction rates.
 As reported in the table, the interaction rate of \pep{} was the only fit component affected by the difference in the CNO values.
 The upper limit on the CNO neutrino interaction rate was obtained with the fit procedure by constraining the ratio of the \pp{} and \pep{} rates, with a Gaussian pull-term, to the values predicted by the SSM.
}
\label{tab:exp_rate}
%\begin{tabular*}{\columnwidth}{@{\extracolsep{\fill}}llll@{}}
 \begin{tabular}{cccc}
 \hline
 \hline
 {Solar $\nu$} & {B16(GS98)-HZ} & {B16(AGSS09)-LZ} & Borexino Results \\
    & \cpd{} & \cpd{}  & \cpd{} \\
\hline
$pp$   & $131.1 \pm 1.4$ & $132.2 \pm 1.4$ & $134 \pm 10 ^{+6}_{-10}$ \\
\hline
$\rm{^7Be}$ & $47.9 \pm 2.8$ & $43.7 \pm 2.5$ & $48.3 \pm 1.1^{+0.4}_{-0.7}$\\
\hline
$pep$  & $2.74 \pm 0.04$ & $2.78 \pm 0.04$& $2.43 \pm 0.36 ^{+0.15}_{-0.22}$(HZ)\\
\hline
&    &    & $2.65 \pm 0.36 ^{+0.15}_{-0.24}$ (LZ) \\
\hline
$\rm{CNO}$ & $4.92 \pm 0.78$ & $3.52 \pm 0.52$& $<8.1$ (95\% C.L.)\\
\hline
\hline
\end{tabular}
\end{table*}
%=====================================

%%%%%%%%%%%% CNO sensitivity %%%%%%%%%%%%
\section{Borexino sensitivity to CNO neutrinos}
\label{sec:CNOStrat:Sens}
\red{In the Borexino Phase-II analysis \cite{ppchainNature}, the signal and background rates were determined using the multivariate fit method.} In detail, as discussed in \cite{BxPhaseI, BxNuSol}, we \red{started} from a background model consisting of a list of unstable isotopes contaminating the scintillator (namely \ce{^{14}C}, $^{85}$Kr, $^{210}$Po, $^{210}$Bi), isotopes of cosmogenic origin ($^{11}$C), and $\gamma$-rays emitted in the regions outside of the sensitive volume and referred to as external backgrounds (originating from decays from $^{208}$Tl, $^{214}$Bi, and $^{40}$K).

We then \red{performed} a multivariate fit procedure of events selected in an optimized wall-less fiducial volume {\color{black} (FV) of about 70\,ton} in the central part of the detector that \red{included} \red{two} observables:
the energy spectra of the events in the region from 0.19 to 2.93\,MeV and their reconstructed radial coordinate distribution~\cite{BxNuSol}. This fitting procedure \red{made} it possible to disentangle the rates of events induced by neutrinos from those induced by backgrounds. At the same time, \red{the fair goodness-of-fit (\(p\)-value of 0.7)} \red{verified} the consistency of the background model.

\red{In this study, we make assumptions of background rates and signal rates based on the multivariate fit analyses of Phase-III data, which starts from 2016 June.} The expected energy distribution of the signals and of the most relevant background components can be seen in Fig.~\ref{fig:BxSpectrum},
while the expected and measured interaction rates of solar neutrinos (Phase-II results) are listed in Table~\ref{tab:exp_rate}\red{, and the expected signal and background rates considered in the present study are listed in Table~\ref{tab:counting}.}

There is an intrinsic difficulty in the measurement of solar neutrinos due to the similarity of the energy spectra of \bi{} electrons and those of electron recoils induced by \pep{} and CNO neutrinos. This feature, which was discussed in \cite{BxPhaseI, BxNuSol}, is clearly visible in Fig.~\ref{fig:BxSpectrum}.
The partial degeneracy of these spectral shapes induces significant correlations among these three components. Only by constraining the rates of two out of three components can one measure the rate of the third precisely.
%As a consequence, the measurements of ${R_{pp}}$, ${R_{\rm Be}}$, and ${R_{pep}}$ ) presented in \cite{ppchainNature} were made possible only by constraining ${R_{\rm CNO}}$ to the values expected from the SSM high (HZ, \cite{GS98}) and low (LZ, \cite{AGSS09}) metallicity models. For the same reason, the limit on ${R_{\rm CNO}}$ (also reported in \cite{ppchainNature}) was achieved by constraining the ratio of ${R_{pp}}$ and ${R_{pep}}$ to the SSM predictions.

However, the measurement of ${R_{\rm CNO}}$ \red{(\textit{i.e.}, the interaction rate of CNO solar neutrinos)} is now possible in Borexino Phase-III \red{as we have identified a strategy (discussed in Sec.~\ref{sec:CNOStrat:Bkg}) to measure or constrain the two main components, \pep{} neutrinos and \bi{} in the scintillator. The thermal stabilization of the detector, achieved during the Phase-III period, is fundamental to measure the rate of \bi{}.} The low internal background levels, great depth of LNGS, passive detector shielding, and active removal (based on the three-fold-coincidence method \cite{BxPhaseI}, \textit{i.e.} TFC cut) of the cosmogenic \ce{^{11}C} background are also crucial.

\begin{figure}[h]
 \centering
 \includegraphics[width=.48\textwidth]{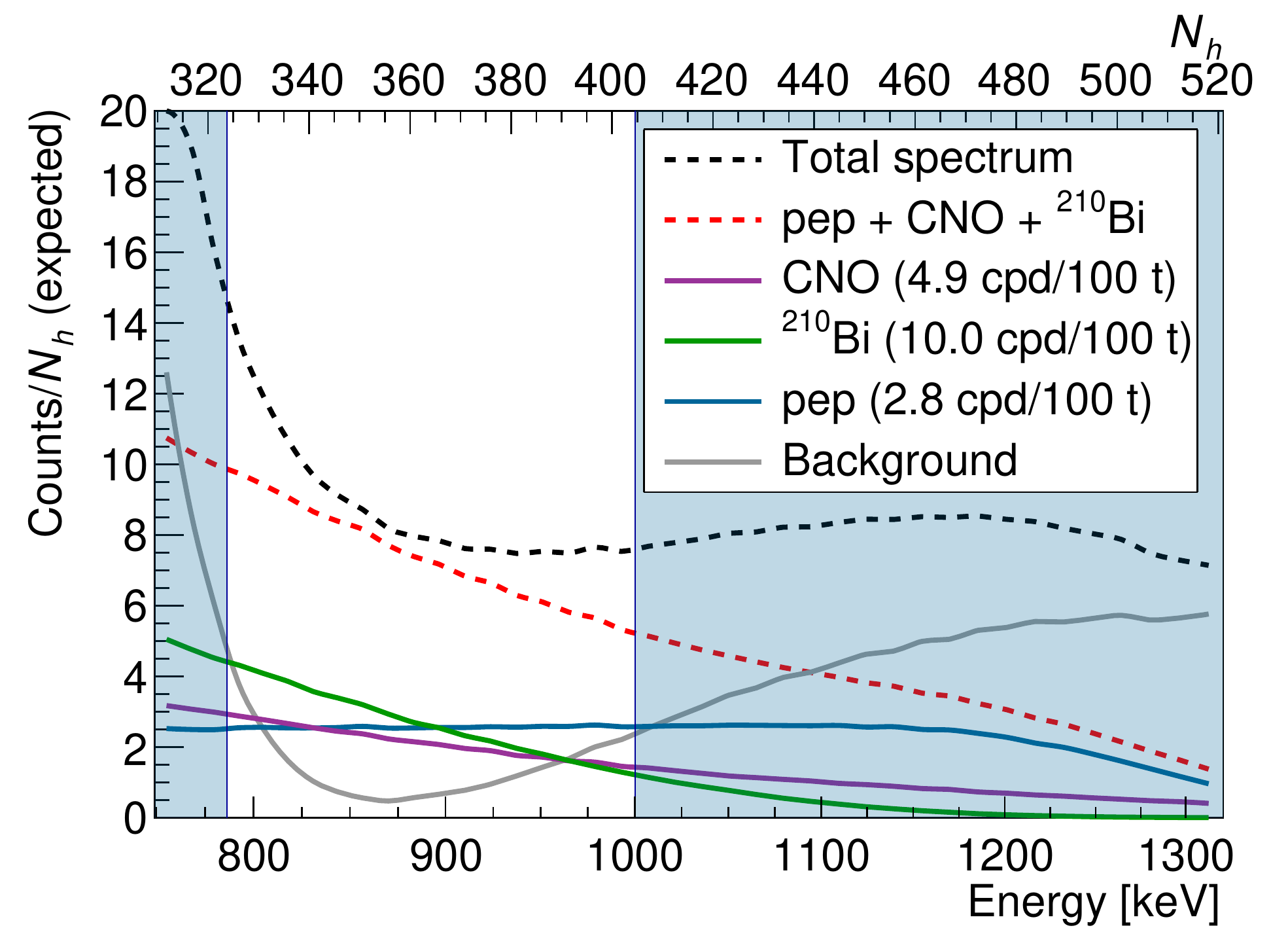}
 \caption
 {
 Expected Borexino event energy distribution of \(pep\) (solid blue line), CNO neutrino (solid purple line), and \bi{} backgrounds (solid green line) in the region of interest (ROI) used in the counting analysis. The sum of three components is shown as the dashed red line. Other backgrounds are marked as the solid gray line, and the total sum is shown as the black dashed line. The assumed rates are listed in Table~\ref{tab:counting}. As in Fig. \ref{fig:BxSpectrum},
 events are selected with the fiducial volume cut and the contribution of \celeven{} is suppressed using the TFC procedure.
 The white region indicates the region of interest.
 }
 \label{fig:ROI}
\end{figure}

Further, we observed an energy region from about \SIrange{0.8}{1}{\mega\electronvolt}, hereinafter referred to as region of interest (ROI), that is dominated by \(R_{\rm Bi}\), \(R_{pep}\), and ${R_{\rm CNO}}$ (see Fig. ~\ref{fig:ROI}). More precisely, in this ROI the background contributions are of the same order as the statistical fluctuations of the total {\color{black}expected} CNO rate. Thus, it is also possible to extract the sensitivity of Borexino to CNO neutrinos through a simple procedure consisting only of determining the background rates and subsequently counting all the events in this energy region.
Section~\ref{sec:counting} first discusses this counting analysis. Then, Sec.~\ref{sec:shape} presents a sensitivity study performed with the multivariate spectral fit. Finally, Sec.~\ref{sec:CNOStrat:Sens:Sigma}-\ref{sec:CNOStrat:Sens:Disc} compare the resulting expected sensitivities and discovery significance {\color{black} from both methods}.

In the following sections, {\color{black} our best estimates of the conditions of the Borexino Phase-III are applied.} The \bi{} rate \(R_\text{Bi}\) is assumed to be \red{10 \cpd{}}. The exposure of \SI{1000}{day}\(\times\)70 ton considers the same fiducial volume as the one applied in the previous solar analysis \cite{ppchainNature} and amounts to {\color{black} $192\,\text{year}\times\text{ton}$} in total. \red{In the counting analysis, only events after the TFC cut are used, and the exposure loss is 36\%. In the multivariate fit analysis, all events are used. In order to facilitate the comparison of the performances of the two methods, we use the same values of the constraint of the \(pep\) neutrino interaction rates as well as the \bi{} contamination of the scintillator. The \pep{} constraint (Sec.~\ref{sec:pep}) is based on the  predictions of the Standard Solar Model, while \bi{} constraint (Sec.~\ref{sec:bi}) is centered in the range considered achievable in Borexino.

We note, that with a high exposure, the intrinsic small difference between the spectral shapes of CNO and \bi{} makes it possible, in principle, to measure the interaction rate of CNO neutrinos without any constraint on \(R_{\rm Bi}\). 
%As shown in Fig. \ref{fig:CNOsigmaShape}, even with the currently assumed exposure, the multivariate fit shows a better performance than the counting analysis when the \bi{} constraint is not very strong. Quantitatively, 
With a spectral analysis similar to that discussed in this paper, we estimate that the statistical significance of about $3\sigma$ can be reached for an exposure of $450\,\text{year}\times\text{ton}$ assuming CNO neutrinos from the HZ SSM. The statistical sensitivity of such analysis improves as the inverse square root of exposure. However, the systematic uncertainties due to the detector response and energy scale modeling must be treated carefully.}

%=====================================
%================================

\subsection{Borexino as a \emph{counting experiment}}
\label{sec:counting}
This approach consists of counting the total number of events in the ROI to obtain the interaction rate of CNO neutrinos as {\color{black}the difference between the total number of events detected in ROI and that of the backgrounds, as evaluated in an independent way.} 

Our key assumptions lie in the availability of a realistic background model and the ability to measure the rate of the residual background components independently (or to label them as negligible). Another assumption is the availability of a sufficiently accurate model of the detector response function, needed to determine the fraction of each spectral component in the ROI.

As shown in Fig. \ref{fig:BxSpectrum} and Fig. \ref{fig:ROI}, there are six contributions to the event rate in the ROI: CNO neutrinos,
\bi{} electrons, \pep{} neutrinos, residual \celeven{} positrons, $^{7}$Be neutrinos, and external gamma backgrounds.
The impact of the relevant parameters like the exposure, the choice of the ROI, and the precision of the estimation of \pep{} neutrino and \bi{} rates can be understood by writing $R_\text{CNO}$ and its statistical uncertainty $ \sigma^{count}_\text{CNO}$ as:
\begin{align}
 R_\text{CNO} &= \frac{1}{\varepsilon_\text{CNO}} \left( r_\text{tot} -
 \varepsilon_\text{Bi} \tilde{R}_\text{Bi}
 - \varepsilon_{pep} \tilde{R}_{pep}-\tilde{r}_\text{o} \right)
 \label{eq:RateCnt}
 \\
 \sigma^{count}_\text{CNO} &= \frac{1}{\varepsilon_\text{CNO}} \left(\sigma_\text{tot}
 \oplus 
 \varepsilon_\text{Bi}\tilde{\sigma}_\text{Bi}
 \oplus 
 \varepsilon_{pep}\tilde{\sigma}_{pep}
 \oplus 
 \tilde{\sigma}_\text{o} \right)
 \label{eq:SigmaCnt}\\
 r_x &= \dfrac{N^\text{ROI}_x}{\mathcal{E}}\:(x=\text{tot},\text{o})\nonumber\\
 \sigma_\text{tot} &= \dfrac{\sqrt{N^\text{ROI}_\text{tot}}}{\mathcal{E}}\nonumber\\
 \sigma_\text{o} &= \varepsilon_\text{Be}\tilde{\sigma}_\text{Be}\oplus
 \varepsilon_\text{C}\tilde{\sigma}_\text{C}\oplus
 \varepsilon_\text{ext Bi}\tilde{\sigma}_\text{ext Bi}\oplus ... \nonumber
 ,
\end{align}
%where \(R_\text{CNO}\) and \(\sigma_\text{CNO}\) are the value and statistical uncertainty of the measured CNO solar neutrino interaction rate,
where \(N^\text{ROI}_x\) (\(x=\)tot(al) or o(thers)) is the number of events in ROI of \(x\), \(\mathcal{E}\) is the exposure,
$\varepsilon$ is the fraction of events for each component falling inside the ROI,
$R$ and \(\sigma\) are the rate and the uncertainty, respectively, of components in the full energy range,
%of the independently determined rates and their precision of the component \(i\) in the full energy range,
and \(r_\text{o}\) is the total event rate of all other components in the ROI. The contributions to \(r_\text{o}\) are either negligible and/or independently determined. \red{The \ce{^7Be} solar neutrino rate is constrained to Phase-II results \cite{ppchainNature}. \ce{^{85}Kr} is determined with the fast coincidence tagging of its minor branch \cite{ppchainNature}. \ce{^{11}C}, external \ce{^{214}Bi}, \ce{^{208}Tl}, and \ce{^{40}K} are determined together with events outside the ROI via the spectral fitting method. \po{} is negligible in the ROI.}
Variables with a tilde mark the values from \red{independent determinations} and the subscripts indicate the spectral component. The symbol \(\oplus\) is defined as \(a\oplus b = \sqrt{a^2+b^2}\).

\red{\begin{table}
\begin{minipage}{\columnwidth}
\centering
\caption{Assumed rates without energy threshold, expected number of events in the ROI after TFC cut, and fraction of events in the ROI (\textit{i.e.} \(\varepsilon\) in Eq.~(\ref{eq:RateCnt})) of each component. The assumed precision of rates of each background is also provided, except that the uncertainty of CNO is calculated from Eq. (\ref{eq:SigmaCnt}). The efficiencies are estimated using an \emph{ab initio} simulation software \cite{bib:BxMC}. The uncertainties of efficiencies are all less than 0.06\%, mainly coming from the uncertainty of the energy scale, so they are neglected. The exposure used here is after TFC cut and is 447 day \(\times\) 100 ton.}
\label{tab:counting}
%\begin{tabular*}{\columnwidth}{@{\extracolsep{\fill}}llll@{}}
 \begin{tabular}{cccc}
 \hline
 \hline
 Component & Rates & Events {\color{black} ($N^{\text{ROI}}$)}  & Efficiencies (\(\varepsilon\)) \\
  & \cpd{} & & \% \\
\hline
Total & & \(697 \pm 26\) & \\
\hline
CNO \(\nu\) & \(4.92\pm1.50\) & \(162\pm49\) & 7.37 \\
\hline
\bi{} & \(10\pm2\) & \(203\pm41\) & 4.55 \\
\hline
\(pep\) \(\nu\) & \(2.74\pm0.04\) & \(195.8\pm2.9\)  & 15.98 \\
\hline
\ce{^7Be} \(\nu\) & \(47.9 \pm 1.3\) & \(61.2\pm1.7\) & 0.29 \\
\ce{^{11}C} & \(1.5 \pm 0.3\) & \(32.9\pm6.6\) & 4.91 \\
ext \ce{^{214}Bi} & \(4\pm0.6\) & \(19.1\pm2.7\) & 1.07\\
ext \ce{^{208}Tl} & \(5\pm0.3\) & \(14.47\pm0.84\) & 0.65\\
ext \ce{^{40}K} & \(1\pm0.7\) & \(6.4\pm4.6\) & 1.42\\
\ce{^{85}Kr} & \(12\pm1.4\) & \(1.23\pm0.14\) & 0.02\\
\ce{^{210}Po} & \(50\pm2.3\) & \(0.06\pm0.01\) & 0.00\\
\hline
\hline
\end{tabular}
\end{minipage}
\end{table}}

\begin{figure}[h]
 \centering
 \includegraphics[width=\columnwidth]{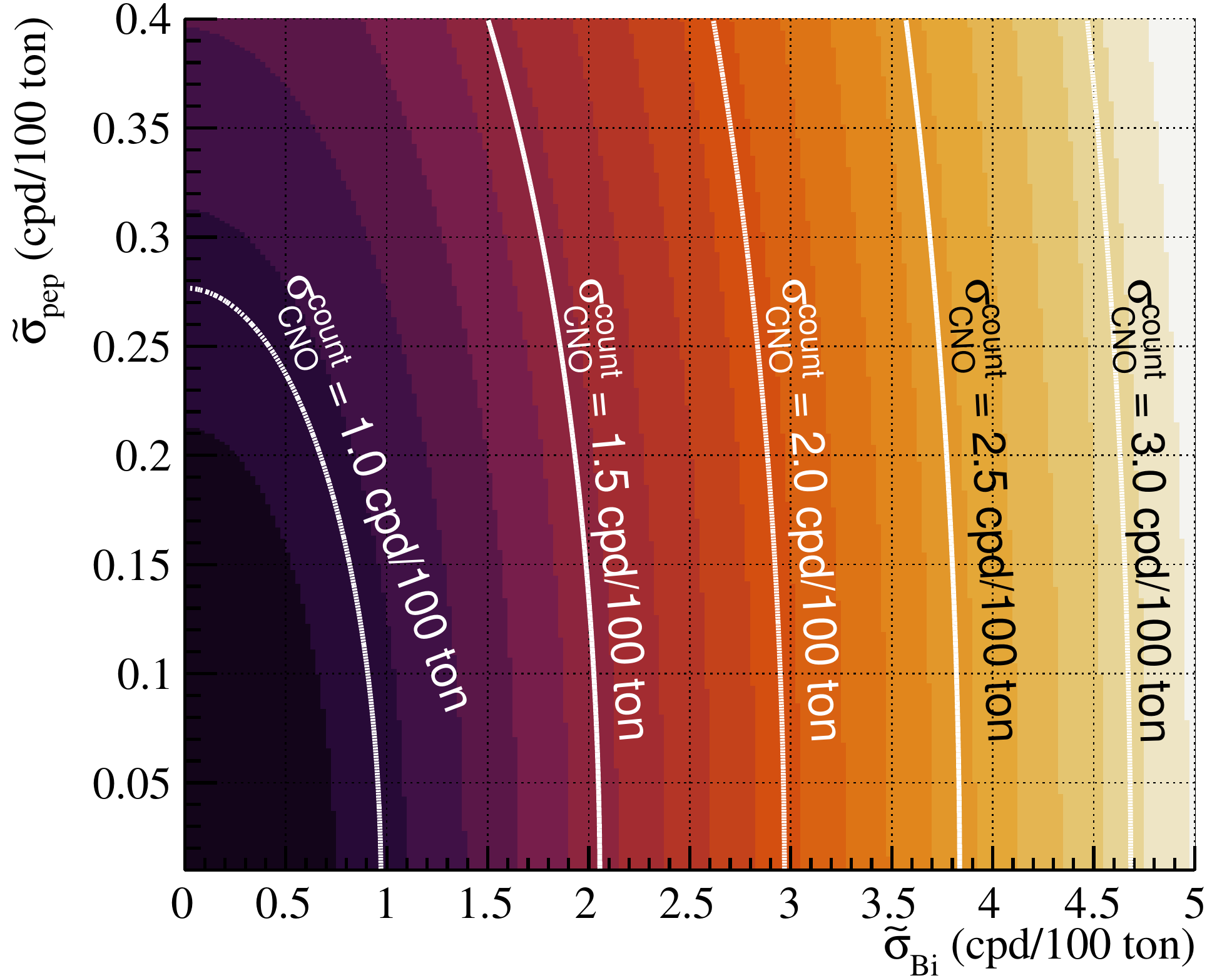}
 \caption
 {
 The uncertainty of the CNO neutrino rate as a function of the uncertainties on \pep{} neutrino and \bi{} rates from the counting analysis (see Eq.~\eqref{eq:SigmaCnt}).
 }
 \label{fig:CNOsigmaCnt}
\end{figure}

\red{The values of the assumed rates, uncertainties, and efficiencies are listed in Table~\ref{tab:counting}. The ratios of the four terms in Eq. (\ref{eq:RateCnt}), from left to right, are \(10\mathpunct{:}3\mathpunct{:}3\mathpunct{:}2\). When \(\tilde{\sigma}_\text{Bi}\) is 1.5 \cpd{} and \(\tilde{\sigma}_{pep}\) is \(1.4\%\cdot \tilde{R}_{pep}\), or 0.04 \cpd{},
the ratios of the four terms in Eq. (\ref{eq:SigmaCnt}), from left to right, are \(9\mathpunct{:}11\mathpunct{:}1\mathpunct{:}3\).}
%Because they are summed up quadratically, the total uncertainty is worsened only when the smallest terms increase to a similar level of the largest.
The largest contribution to $ \sigma^{count}_\text{CNO}$ in Eq.~(\ref{eq:SigmaCnt}) is the \bi{} term
unless the uncertainty on the \bi{} constraint is lower than 1.3 \cpd{} \red{when \(\sigma_\text{tot}=\varepsilon_\text{Bi} \tilde{R}_\text{Bi}\)}, in which case $ \sigma^{count}_\text{CNO}$ is limited by a statistical error.
The resulting uncertainty on the CNO neutrino interaction rate as a function \(\tilde{\sigma}_\text{Bi}\) and \(\tilde{\sigma}_{pep}\) can be seen in Fig.~\ref{fig:CNOsigmaCnt}.

% This belongs elsewhere!
%Because $R_\text{CNO}$ is predicted to be $3.52\,(11\%)$ and $4.91\,(10\%)$ \cpd{} (see Table \ref {tab:exp_rate}) according to the LZ and HZ models, respectively, the total uncertainty of the measured CNO rate needs to be less than or around 1.5 \cpd{} to be able to claim a \(3 \sigma\) evidence of CNO solar neutrinos detection.

%================================================

\subsection{Multivariate fit analysis}
\label{sec:shape}
We performed a study of the sensitivity of Borexino to CNO neutrinos by simulating tens of thousands of pseudo-experiments and fitting the pseudo-data with the full multivariate procedure as the one used for real data \cite{ppchainNature}.
This approach requires maximizing a two dimensional likelihood function built using \red{two observables (energy estimator and radial position)}.
We assumed, as throughout Sec.~\ref{sec:CNOStrat:Sens}, that ${R_{pep}}$ and \(R_{\rm Bi}\) can be independently estimated. The constraints on ${R_{pep}}$ and \(R_{\rm Bi}\) were implemented by introducing two Gaussian pull terms in the likelihood function. The centroid of the pull terms are varied according to the uncertainty of the pull. The uncertainties of the two constraints are the most important parameters when determining the sensitivity.

Each simulated dataset was fitted assuming the same detector response function as that used in the simulation. The 
distribution of the best-fit results for the CNO interaction rate was used to assess the expected statistical uncertainty and discovery significance. The results were obtained using the multivariate fitting tools \texttt{GooStats} \cite{Ding2018} and \texttt{m-stats} \cite{mstats}.

Similarly to the data analysis of \cite{ppchainNature}, the study was conducted using both Monte Carlo and analytical methods that are complementary to each other.
In the Monte Carlo approach the detector response was modeled using an \textit{ab initio} simulation software \cite{bib:BxMC} that takes into account realistic and microscopic descriptions of the energy deposition, scintillation light generation and propagation, electronics response, and energy and position reconstruction.
In the analytical approach, the detector response was modeled using analytical functions \cite{BxNuSol}. The analytical procedure provides an independent way of evaluating the impact of the systematic uncertainty related to detector response parameters (\textit{e.g.}, the light yield). The two approaches gave consistent results for the expected uncertainty and discovery significance.

\subsection{Uncertainty of the CNO rate}
\label{sec:CNOStrat:Sens:Sigma}

\begin{figure}[hbt]
 \centering
 \includegraphics[width=0.45\textwidth]
 {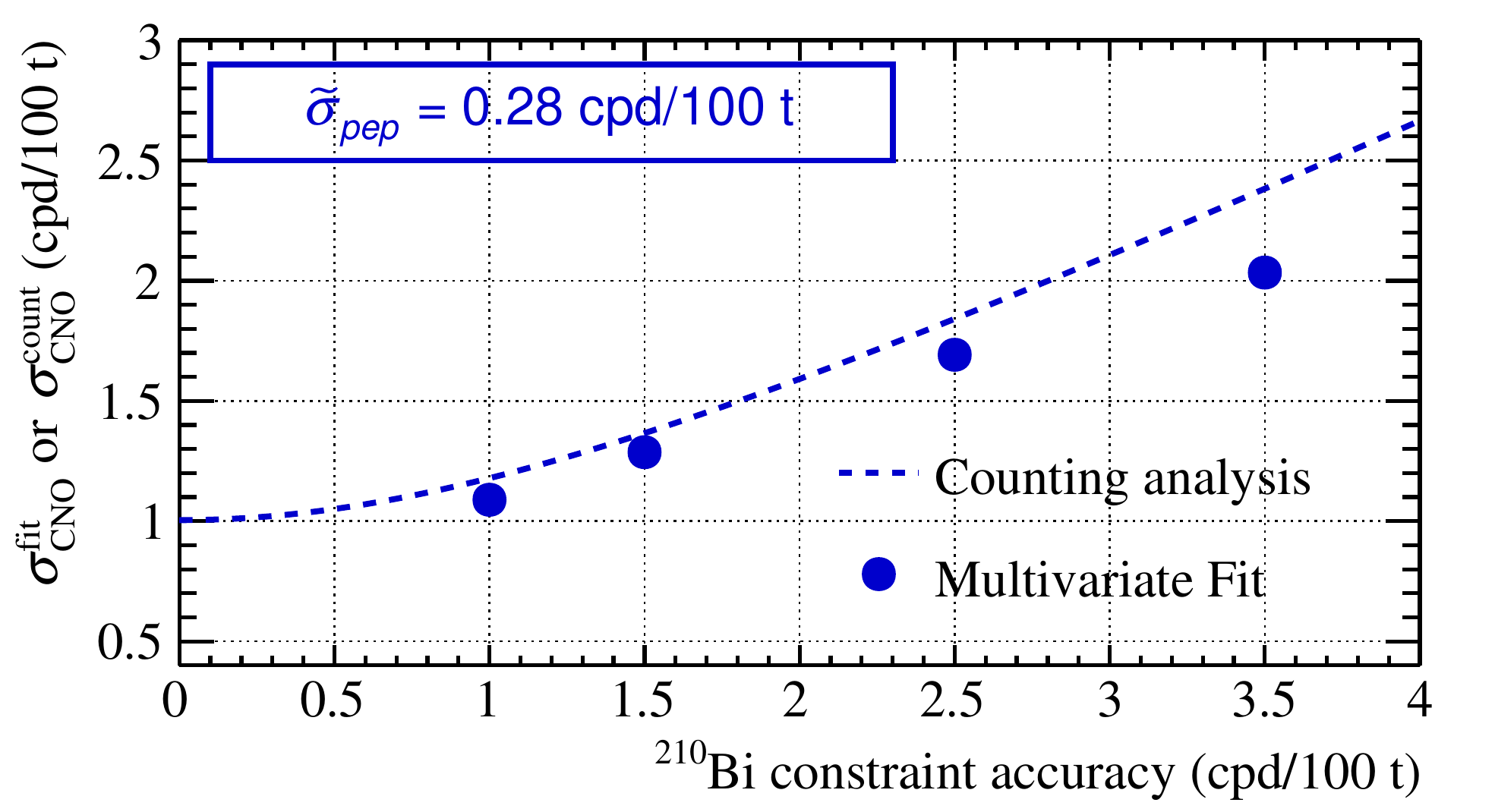}
 \includegraphics[width=0.45\textwidth]
 {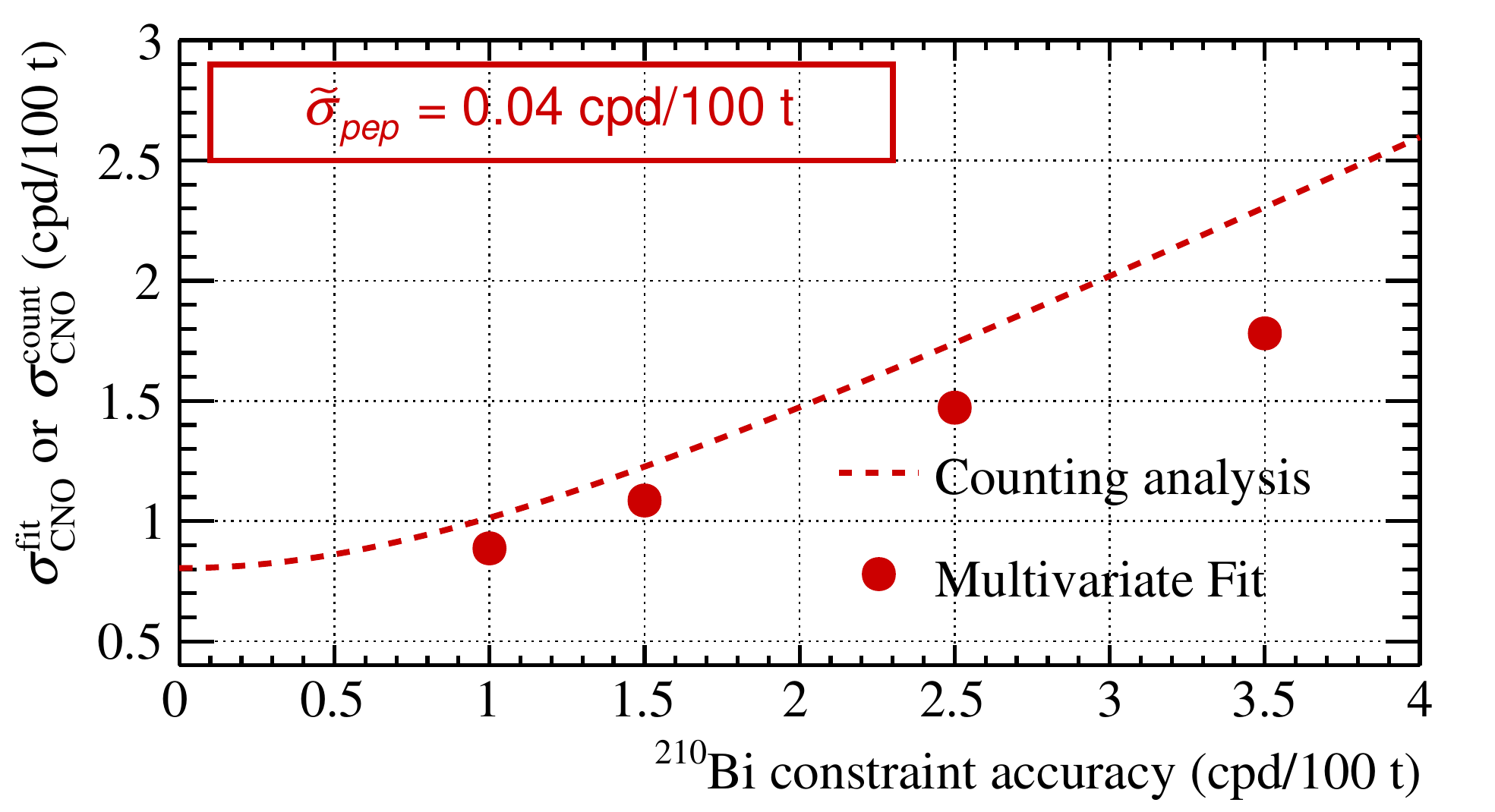}
 \caption[Expected uncertainty on the $\nu(CNO)$ rate from 
 a rate + shape analysis]
 {Comparison between \(\sigma_{\rm CNO}^{fit}\) (in solid dots) and \(\sigma_{\rm CNO}^{count}\) (in dashed lines)
 for two values of \(\tilde{\sigma}_{pep}\) and multiple \(\tilde{\sigma}_\text{Bi}\).}
 \label{fig:CNOsigmaShape}
\end{figure}

In this section the uncertainty of the CNO rate {\color{black} resulting from} both the counting and multivariate fit analyses is discussed. In the framework of the multivariate analysis, the distribution of the values of $ R_{\rm {CNO}} $ resulting from fits performed with fixed external constraints is built. The width of this distribution is used to estimate $\sigma^{fit}_\text{CNO}$.
Figure~\ref{fig:CNOsigmaShape} shows $\sigma^{fit}_\text {CNO}$ as a function of the \bi{} \red{constraint uncertainty} for two different values of \(\tilde{\sigma}_{pep}\) as well as the uncertainty $\sigma^{count}_\text{CNO}$ resulting from the counting analysis presented in Sec. \ref{sec:counting}.

%Figure~\ref{fig:CNOsigmaShape} shows that the results of two analyses are fully compatible when \(\tilde{\sigma}_{\rm Bi}\lesssim 1.5\) \cpd{}.
%One can conclude that the inclusion of all of the additional information used by the multivariate fit method only marginally helps to constrain the ratios \(R_\text{CNO}/R_\text{Bi}\) and \(R_\text{CNO}/R_{pep}\) {\color{blue} when the constraint on $^{210}$Bi is strong}.
% {\color{blue} lets be careful, we published 5 sigma with fit and 3.5 sigma with counting. Whate are the new numbers after taking in the counting analyiss the error sigma tot only the square root of the counts in ROI without the special treatment that was done to produce this plot?}

In the counting analysis, the rates of \bi{} and $pep$ 
are fixed by the external constraints. Thus, according to Eq.~\eqref{eq:RateCnt}, a bias on the constraints will transfer linearly to the 
reconstructed CNO rate. 
The situation becomes slightly more complicated when a full spectral analysis, combined with constraints on the background, is performed.
Indeed, in the case of a bias in the external constraints, the impact on the value of the extracted CNO rate is mitigated by the tension between the shape information and the biased values. 

The systematic uncertainty of the CNO rate is dominated by the bias of the \bi{} accuracy, (see Sec. \ref{sec:bi}), followed by the accuracy on the light yield.

%{\color{blue}Livia; I suggest to move the LS purification section here with some modifications}

\subsection{Expected discovery significance to CNO neutrinos}
\label{sec:CNOStrat:Sens:Disc}
A frequentist hypothesis test was performed to assess the expected discovery significance to CNO neutrinos.
%Using a profile likelihood test statistics $q_0$ \cite{CowanProfileL}.
In the search for the CNO signal, two hypotheses are considered:
the \emph{null} hypothesis, $H_0$, where no CNO signal is present; and the \emph{alternative} hypothesis, $H_1$, that includes
the presence of a CNO signal in addition to the background. 
We used the profile likelihood ratio $q_0$ \red{as the test statistic \cite{CowanProfileL}:
\begin{align}
    q_0 = -2 \ln\dfrac{L(H_0)}{L(H_1)},
\end{align}
where \(L(H_0)\) and \(L(H_1)\) are the the maximum values of the likelihood against $H_0$ and $H_1$.}
The significance of the signal \red{of a measurement} is quantified by evaluating 
the compatibility of the observed data with the null 
hypothesis $H_0$ \red{and represented by the \(p\)-value. For a specific measurement with a test statistic value \(q_{0,\text{obs}}\), its \(p\)-value is the probability that the test statistic is above \(q_{0,\text{obs}}\) under \(H_0\). In quantifying the expected sensitivity of \(H_1\) to discovery of a signal, the \emph{median discovery significance} is used. The \emph{median discovery significance} of \(H_1\) is the \(p\)-value of the median value of the test statistic under \(H_1\).}
This requires the \red{probability density functions (hereinafter as PDFs)} of $q_0$ for both the null 
hypothesis \(f(q_0|H_0)\) and the alternative hypotheses \(f(q_0|H_1=\text{HZ or LZ})\). 

The PDFs of the test statistics $q_0$ were obtained by
analyzing {\color{black} the two sets of the simulated datasets (with and without the CNO injected}), with both the simplified counting analysis and full multivariate fit. {\color{black} We generate 20,000 pseudo-experiments to construct the PDFs. When considering the null hypothesis, \(q_0\) is distributed according to \(1/2\delta(q_0)+\chi^2_1(q_0)\), where \(\delta(q_0)\) is the Dirac-Delta function, and \(\chi^2_1\) is the PDF of the chi-square distribution with one degree of freedom. For \(H_1\) and in the case of uncertainties of 0.04 \cpd{} and \red{1.5} \cpd\ on \pep{} neutrinos and \bi{}, respectively, the median value of the test statistic is 8.0 for HZ SSMs and 4.1 for LZ SSMs.}

\begin{figure}[htb]
 \centering
 \includegraphics[width=\columnwidth]
 {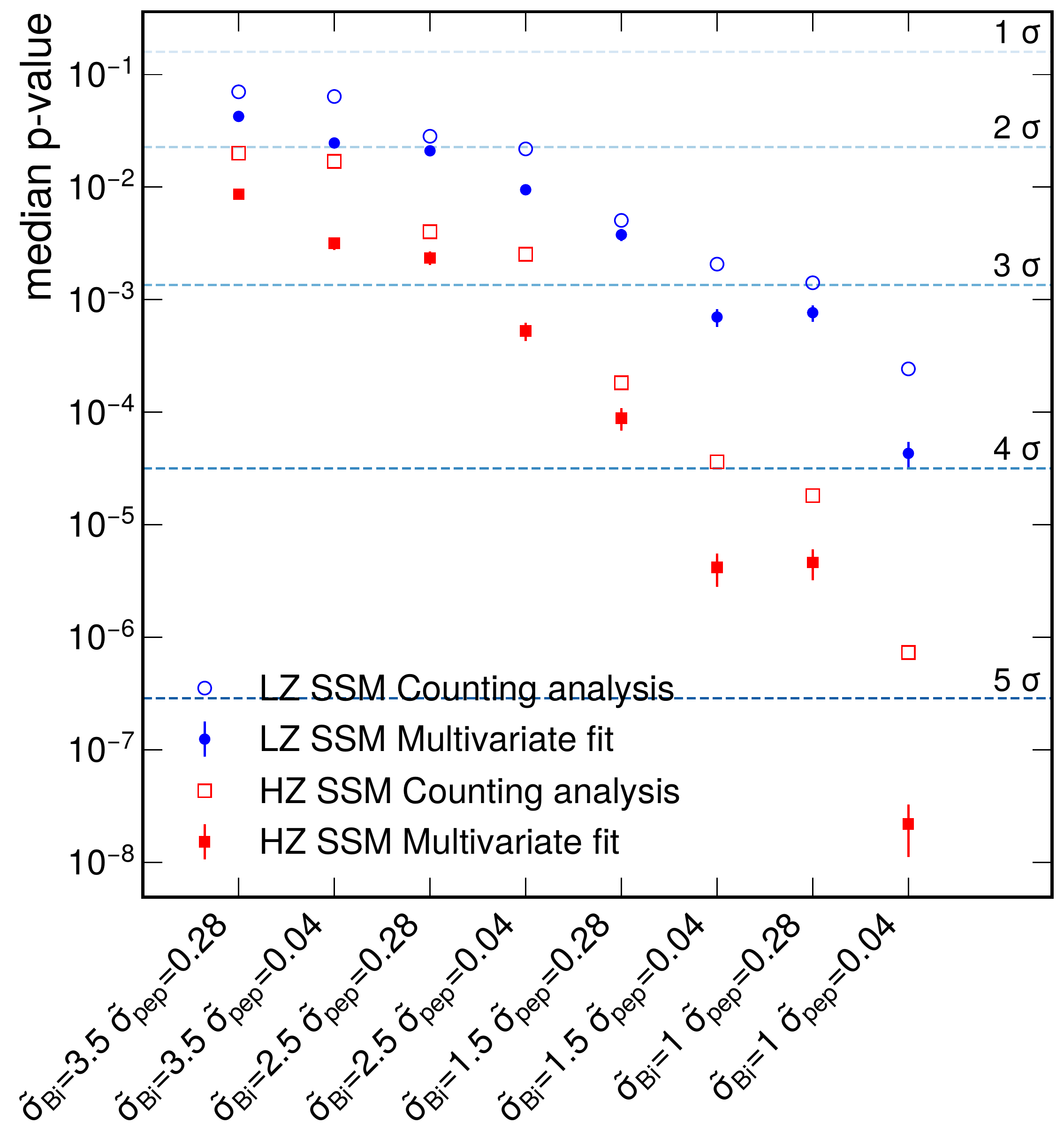}
 \caption[Median discovery significance of a CNO neutrino signal 
 for different accuracies in the determination of the 
 background rate.]
 { 
 Median discovery significance of a CNO neutrino signal
 as predicted by the HZ (in red) and LZ (in blue) SSM 
 for different precision of the constraints on the
 background rates. The results of the counting analysis are indicated
 by empty markers and those obtained with the full multivariate fit 
 by filled markers. The uncertainty \(\tilde{\sigma}_{\rm Bi}\) and \(\tilde{\sigma}_{pep}\) are in \cpd{}.}
 \label{fig:pval}
\end{figure}

The expected discovery significance depends on the strength of the signal 
and on the precision of the external constraints:
a higher rate of CNO neutrinos, as well as 
stronger constraints, results in a higher sensitivity. For the CNO rate predicted by the HZ SSM and in the case of uncertainties of 0.04 \cpd{} and \red{1.5} \cpd\ on \pep{} neutrinos and \bi{}, respectively, the median discovery significance to CNO neutrinos is $4.5\sigma$. \red{In case of the LZ SSM, it is $3.2\sigma$. Results of the expected discovery significance under other conditions are shown in Figure~\ref{fig:pval}.}

The difference between the median $p$-value obtained from the simple counting 
analysis and the one from the multivariate analysis
gets smaller as the precision of the constraints increase.
This effect is coherent with the results on the statistical
uncertainty of CNO presented in Sec.~\ref{sec:CNOStrat:Sens:Sigma}, 
indicating that the impact of the spectral shape information is 
larger when the constraints are relatively weak (high uncertainty), while it becomes 
negligible when the background constraints get more stringent (low uncertainty).

%============================
\subsection{Impact of an upper limit on \bi{} rate}
\label{sec:CNOStrat:Sens:UpLim}

The method discussed in Sec. \ref{sec:bi}, used to obtain an independent measurement of the \bi{} rate, can only provide an upper limit on the \bi{} background under less stringent assumptions, that an additional contribution from migrated (diffusion and/or convection) \po{} (see Sec. \ref{sec:bi}) is present. The presence of migrated \po{} leads to a positive bias of the estimation of the \bi{} rate.

\begin{figure}
 \includegraphics[width=\columnwidth] 
 {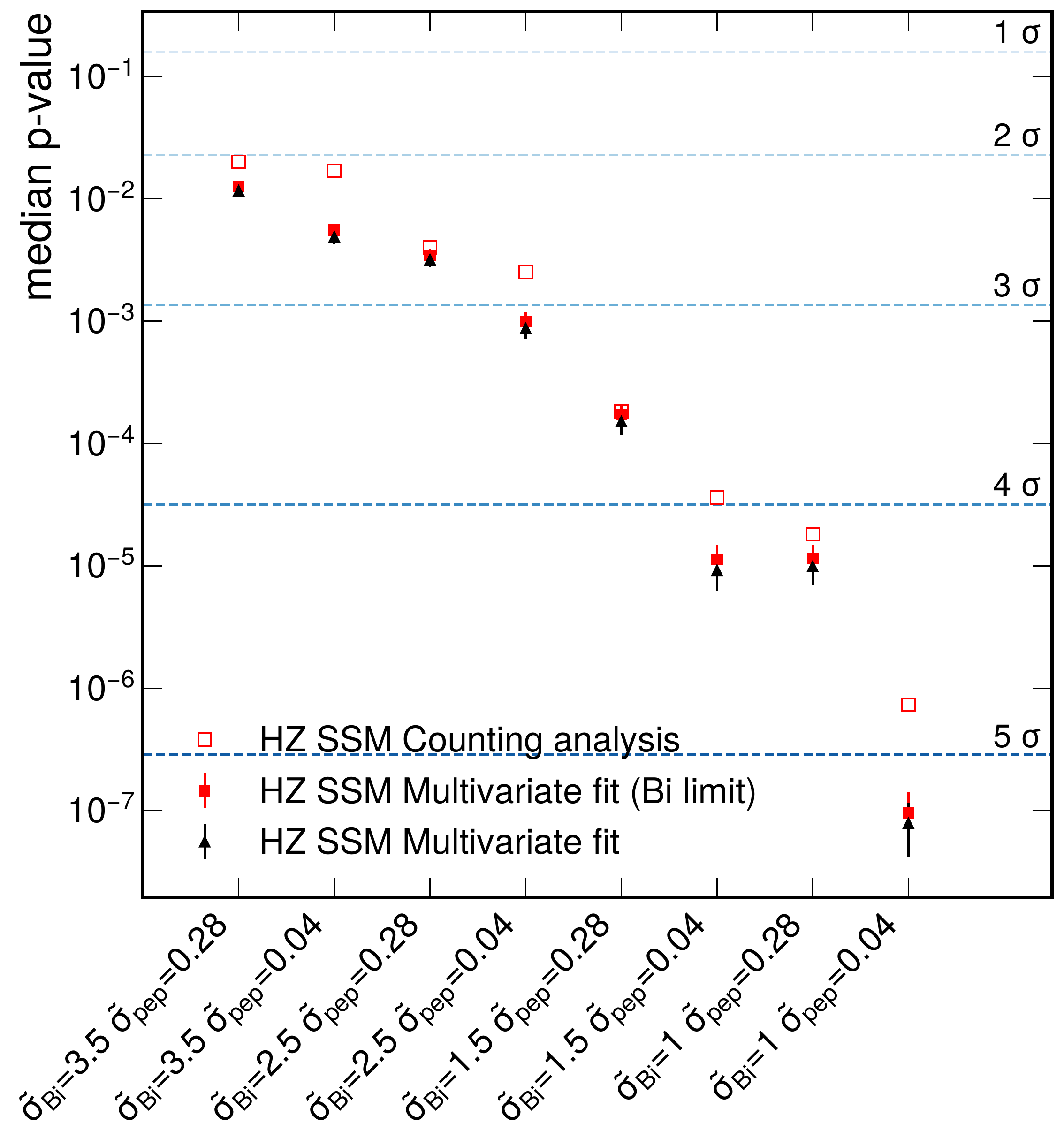}
 \caption[Discovery power in case of upper limit on the 
 \bi{} rate]
 {
 Median discovery significance for the HZ hypothesis on the CNO rate when constraining the \bi{} rate with an 
 upper limit and a symmetric Gaussian penalty. The uncertainty \(\tilde{\sigma}_{\rm Bi}\) and \(\tilde{\sigma}_{pep}\) are in \cpd{}.}
 \label{fig:disc_uplm}
\end{figure}

In this case, the constraint is implemented in the fitting procedure as a one-sided Gaussian penalty\red{:
\begin{align}
    \text{pull term}& = \left(\dfrac{R_\text{Bi}+R_\text{mig Po}-\mu}{\sigma}\right)^2 \\
    R_\text{mig Po}&>0, \nonumber
\end{align}
}
where \(R_\text{mig Po}\) is the rate of the migrated \po{} as discussed in Sec.~\ref{sec:bi}, and \(\mu\) and \(\sigma\) are the centroid and uncertainty of the \bi{} pull term. In so doing, the upper limit on the \bi{} rate is equivalent to a lower limit on the CNO rate.
Therefore, the expected discovery significance for CNO is the same as would be obtained using a two-sided {\color{black} Gaussian} constraint on the \bi{} rate. 
This was confirmed by performing a hypothesis test analogous to that described in Sec.~\ref{sec:CNOStrat:Sens:Disc}.
The results for varying widths ($\tilde{\sigma}_{\text{Bi}}$) on the \bi{} penalty are shown 
in Fig.~\ref{fig:disc_uplm}. The expected discovery significance obtained
from the counting analysis and from applying a symmetric constraint 
are also displayed for comparison.
As one can see, allowing an additional contribution from the migrated \po{} barely decreases the expected discovery significance for CNO.

\begin{figure*}[h]
 \centering
 \subfloat[free]
 {
 \includegraphics[width=.32\textwidth]
 {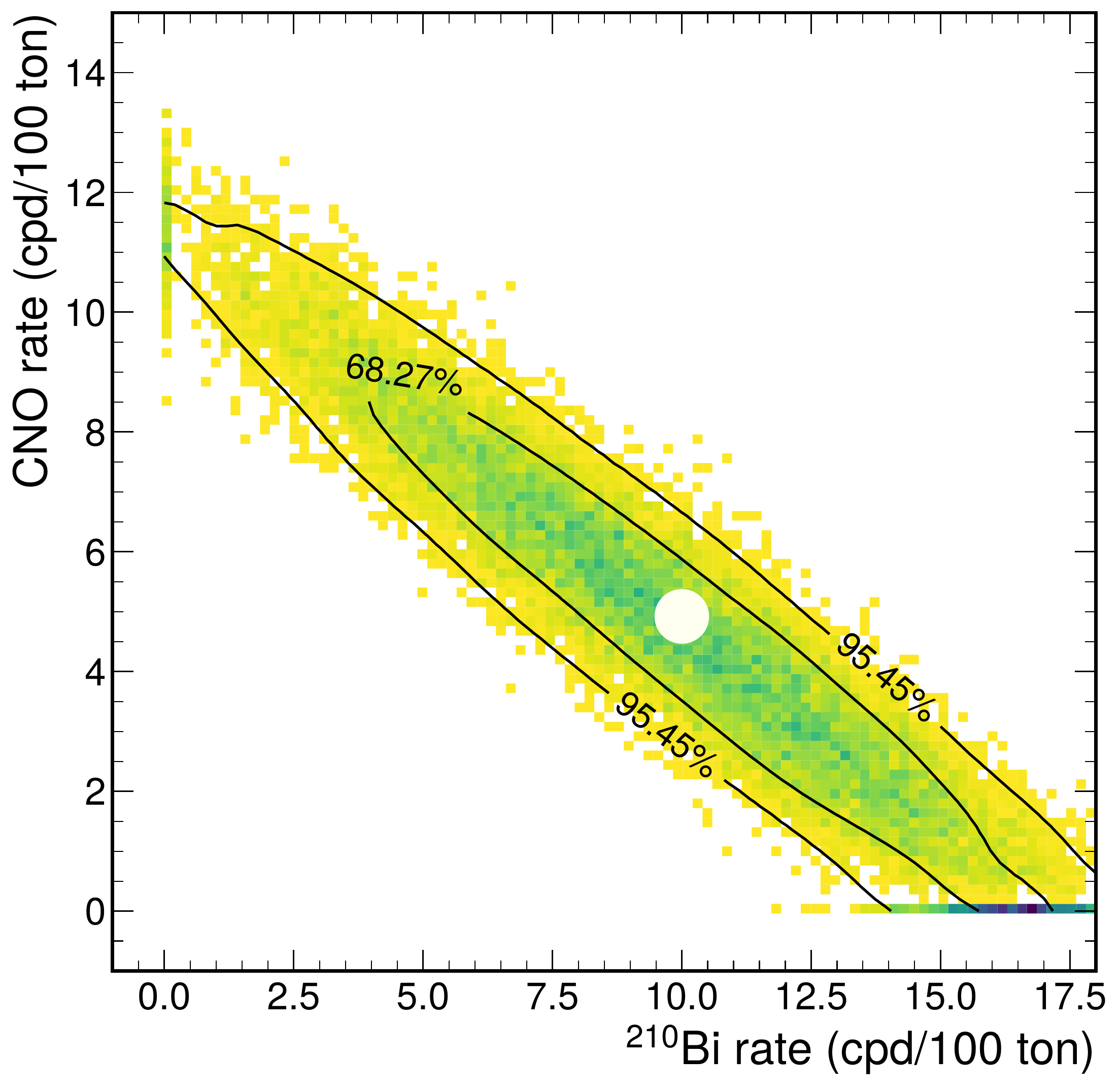}
 \label{fig:BiFree}
 }
 \subfloat[symmetric Gaussian penalty]
 {
 \includegraphics[width=.32\textwidth]
 {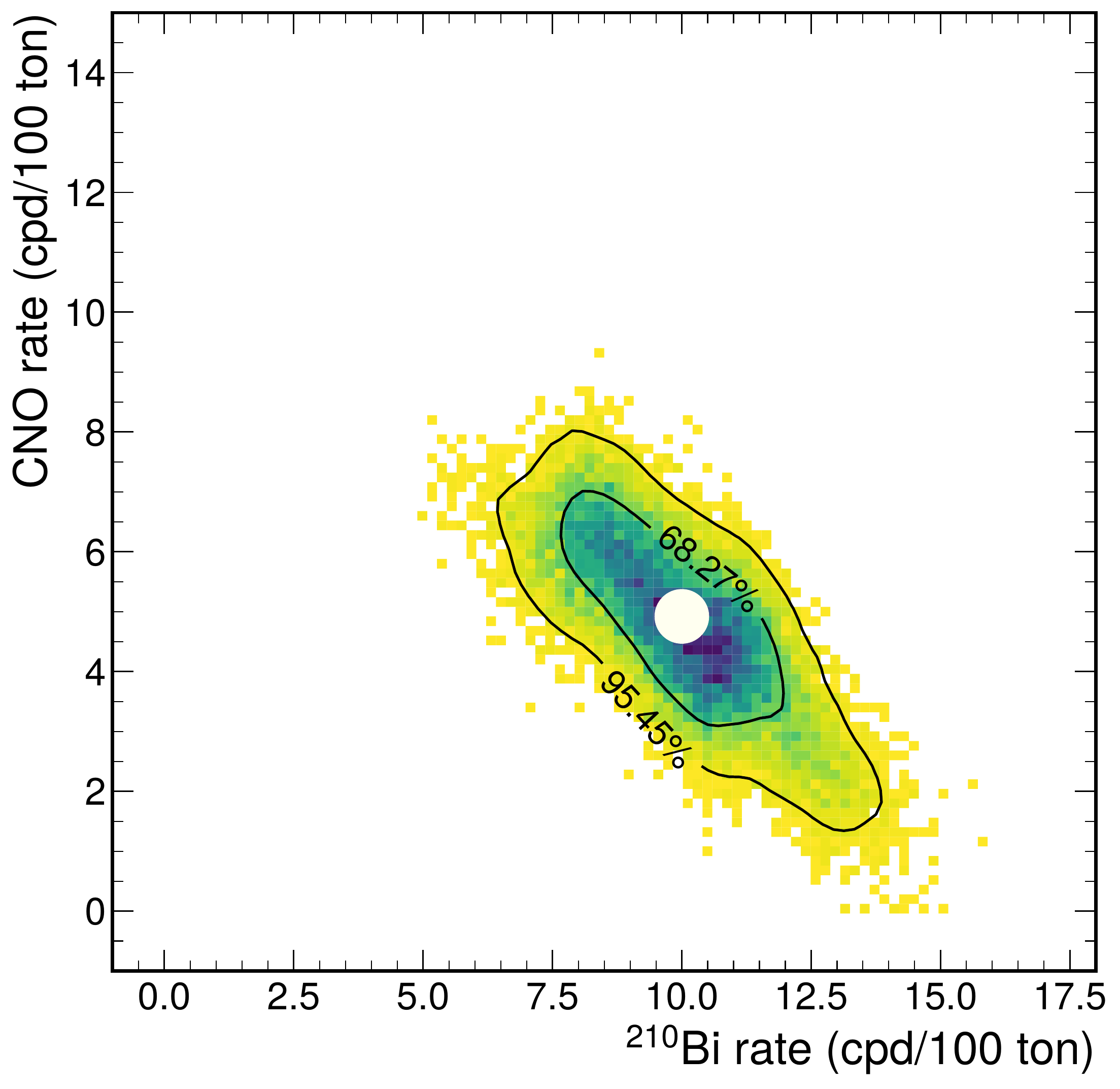}
 \label{fig:BiMeas}
 }
 \subfloat[upper limit]
 {
 \includegraphics[width=.32\textwidth]
 {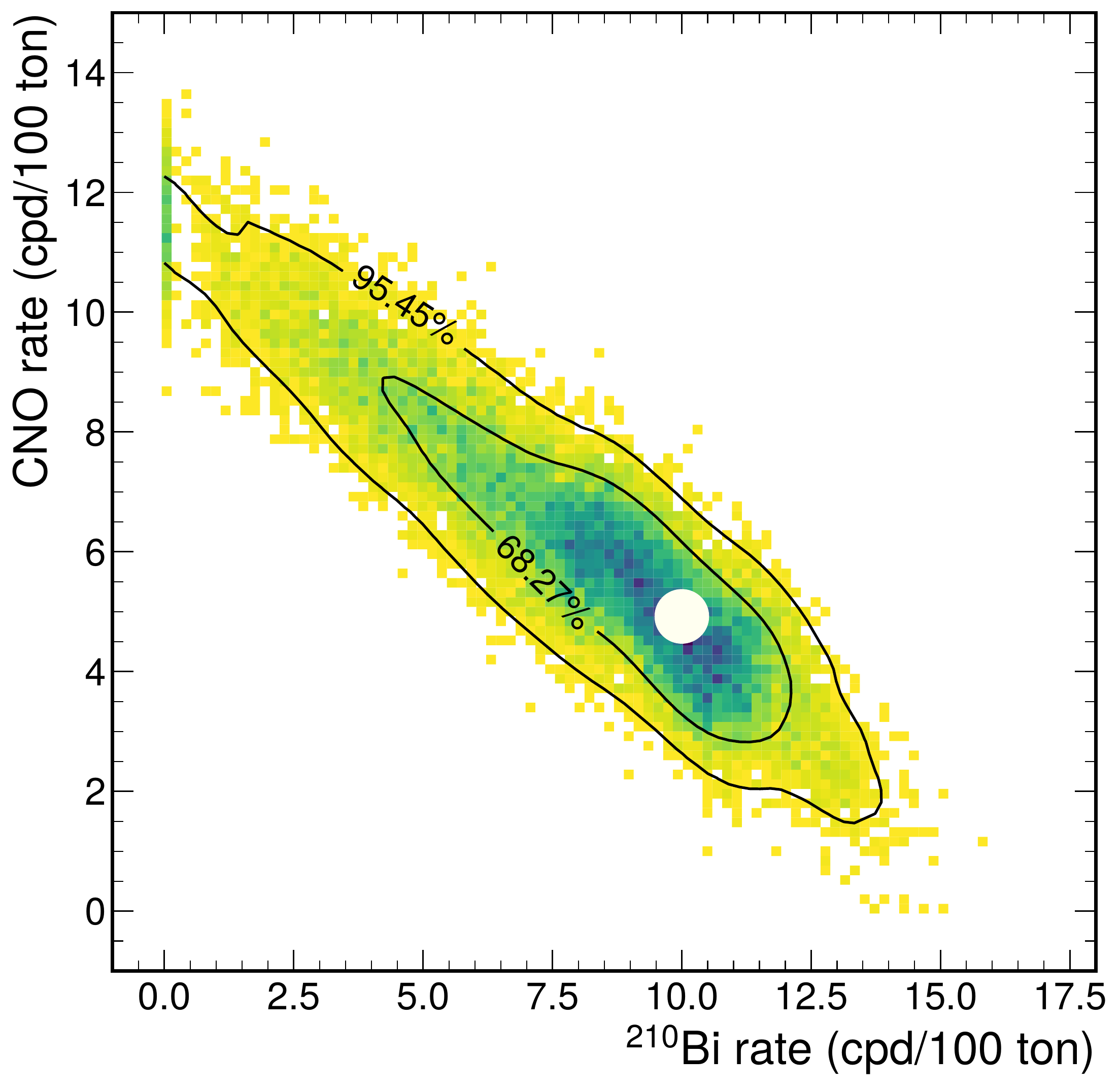}
 \label{fig:BiLimit}
 }
 \caption
 [ Distribution of the best fit estimates in the \bi{}--CNO
 space for the \bi{} rate free, constrained, or upward limited.
 ]
 {
 Distribution of the best fit results of \bi{} and CNO neutrinos interaction rates 
 obtained fitting thousands of simulated datasets
 applying a $10\%$ constraint 
 on the \pep{} rate and (a) keeping \bi{} free, (b) constraining \bi{} with a symmetric Gaussian penalty at a precision of 1.5 \cpd{}, and (c) imposing an upper limit on \bi{}. The injected rates of \bi{} and CNO neutrinos, shown in the figures as white dots, are 10 and 4.9 \cpd{}, respectively.
 }
 \label{fig:BiLimitSeries}
\end{figure*}

Therefore, an upper limit on the \bi{} rate may be sufficient to 
claim detection of CNO neutrinos, but it would not allow a precise measurement of the CNO interaction rate required to solve the solar metallicity problem.
Since there is no lower limit on the \bi{} rate, the upper limit of the CNO neutrino rate 
will not be stringent {\color{black} and the central CNO value might have a negative bias}. 
Figure~\ref{fig:BiLimitSeries} shows the distribution of 
the best fit estimates, obtained from the fit of simulated datasets
when applying a constraint to the \pep{} neutrino rate and 
leaving the \bi{} rate free (Fig.~\ref{fig:BiFree}), 
constraining the \bi{} rate with a symmetric Gaussian penalty 
(Fig.~\ref{fig:BiMeas}), and applying the upper limit 
described in this section (Fig.~\ref{fig:BiLimit}).

The expected confidence intervals in Fig.~\ref{fig:BiLimit} 
are asymmetric: the upper limit of the confidence interval is 
similar to the one obtained when leaving the \bi{} rate free, 
while its lower limit resembles the one resulting when 
the \bi{} rate is constrained.

%=============================

%+++++++++++++++++++++++++++++++++++++++++++++++
\section{Strategy for establishing background assessment }
\label{sec:CNOStrat:Bkg}
This section discusses the strategy for setting constraints on the \pep{} neutrinos and \bi{} rates independently of both the counting analysis and the multivariate fit procedure.

\subsection{$pep$ constraint}
\label{sec:pep}
The electron capture reaction $p+e^{-}+p \to d +\nu_{e}$, which generates 
\pep{} neutrinos in the Sun, is linked to the $\beta^+$ decay 
process $p+p\to d + e^{+} + \nu_{e}$, which is responsible for the \pp{} neutrino production, by well-known nuclear physics.
Since the two processes depend on the same allowed nuclear matrix element, 
the ratio between their rates is determined by the available reaction phase spaces and by 
the electron density $n_{e}$ of the solar plasma only.
{\color{black}This ratio} was calculated in \cite{Bahcall:1968wz, Bahcall:1990tb}, 
and the effect of radiative corrections subsequently discussed in \cite{Kurylov:2002vj} (see, for example, \cite{Adelberger:2010qa} for a review). 
It can be determined with $\sim1\%$ precision \cite{Adelberger:2010qa} for the conditions of the solar interior 
and is mildly dependent on the properties of the solar plasma, 
roughly proportional to $T_c^{-1/2} n_{e}$ (where $T_c$ is the temperature of the core of the Sun).
As a consequence, the ratio $ \Phi_{\rm pep}/\Phi_{\rm pp} $ 
%$\xi\equiv \Phi_{\rm pep}/\Phi_{\rm pp} $ 
between the {\em pep} and {\em pp} neutrino fluxes is a robust prediction of SSMs, and 
it can be used to improve the sensitivity 
to the CNO neutrino signal.
With this approach, Borexino's direct observation of the \pp{} neutrino flux 
\cite{ppchainNature} can be translated into a $\sim10\%$ determination of the 
\pep{}-neutrino component, motivating the analysis performed 
in Sec.~\ref{sec:CNOStrat:Sens}, which assumes $\tilde{\sigma}_{\pep{}} = 0.28$ \cpd{}. 

The precision of the \pp{} and \pep{} neutrino flux determination can be further 
improved by performing a global analysis \cite{NuPars,Bergstrom:2016cbh} on all neutrino experiment results applying the so-called solar luminosity constraint \cite{Bahcall:2001pf,Castellani:1996cm,Vissani:2018vxe}.
In this way, it was shown in \cite{Bergstrom:2016cbh}
that the \pep{} neutrino {\color{black} flux is constrained with $\sim 1\%$ precision of 
by solar neutrino data. By taking into account an additional $\sim 1\%$ uncertainty in neutrino survival probability considering the matter effects in the Sun and the uncertainty of oscillation parameters, the event rate $R_{pep}$
can be constrained} 
with a precision of 
$\sim1.4\%$, motivating the assumption 
$\tilde{\sigma}_{\rm pep} = 0.04$ \cpd{} 
considered in Sec. \ref{sec:CNOStrat:Sens}. Because the contribution of CNO neutrinos to the solar luminosity is around 1\%, neglecting the dependence of this \pep{} constraint on the assumed \(R_{\rm CNO}\) has almost no effects on the significance to CNO neutrinos, as shown in Fig. \ref{fig:CNOsigmaCnt}.

\subsection{\bi{} constraint}
\label{sec:bi}

\bi{} is a \(\beta\)-emitting daughter of \pb{} with a mean lifetime of 7.23 days and a \(Q\) value of 1160 keV. Since the lifetime of \bi{} is small, it must be supported by its long-lasting parent nucleus \pb{} ($\tau = 32.2$ years)
to maintain a constant decay rate. \pb{} is a part of the \ce{^{238}U} chain, and due to its low \(Q\) value {\color{black} of 64 keV} it does not represent a background in this analysis.
{\color{black} The $^{210}$Pb and $^{210}$Bi are part of the of the \ce{^{238}U} chain.} The \red{concentration of \ce{^{238}U} is estimated to be less than \(9.4\times10^{-20}\) g/g (95\% C.L.)} by
the absence of $^{226}$Ra and $^{222}$Rn, easily detectable through the fast $^{214}$Bi-$^{214}$Po coincidence following their decays {\color{black} in the part of the chain above $^{210}$Pb.
 We note, that the secular equilibrium of the chain is broken at the level of $^{210}$Pb. This means, that while the background from the $^{238}$U chain down to $^{210}$Pb is negligible, the background from the chain below $^{210}$Pb is important.}
The purification campaign in 2011 \cite{Purification2012}, {\color{black} that marked the start of the Borexino Phase-II, has} significantly reduced in short order the initial \pb{} contamination, leading to a residual rate in the range of a few tens of \cpd{}.

The strategy to independently determine the \bi{} rate using the decay rate of its decay product, \po{}, was suggested in \cite{PoBipaper}. The isotope \po{}, with a mean lifetime of 199.6 days, ends the chain by decaying into stable $^{206}$Pb thereby emitting a 5305 keV $\alpha$ particle (visible energy around 400 keV electron equivalent):
\begin{equation*}
 ^{210}\text{Pb} \xrightarrow[32.2 y] {\beta}
{^{210}\text{Bi}} \xrightarrow[7.23 d] {\beta}
{ ^{210}\text{Po}} \xrightarrow[199.6 d]{\alpha}
{ ^{206}\text{Pb}}\text{(stable)}.
\end{equation*}

When the above chain is in equilibrium, the \bi{} rate is equal to that of \po{}.
The \po{} rate can be precisely determined in Borexino because \po{} emits \(\alpha\) particles that can be identified event-by-event via pulse-shape discrimination. Using a pulse-shape analysis based on a multi-layer perceptron discriminator~\cite{SeasonalModulation,Agostini2020q},
\po{} events in Borexino are identified with an efficiency very close to 1.

Unfortunately,
the measured \po{} rate is not only due to \bi{} decays, but consists of three components:
\begin{enumerate}
\item \textbf{unsupported \po{}}: the residual \po{} left over by the water extraction phase of the scintillator purification campaign and not linked to local \bi{}.
\item \textbf{migrated \po{}}: originally produced {\color{black} from the \pb{} lead decays } on the inner surface of the nylon vessel holding the liquid scintillator and brought into the fiducial volume by convective and diffusive motions of the liquid scintillator.
\item \textbf{supported \po{}}: in secular equilibrium with local \bi{} present in the liquid scintillator.
\end{enumerate}

The \bi{} rate is equal to the supported \po{} rate and thus less than the total \po{} rate in presence of the unsupported and migration terms. The rate of unsupported \po{}, following the law of radioactive decay, gets asymptotically closer to zero over time \red{with a mean lifetime of 199.6 days.} The migrated \po{} is the most intricate contribution to handle. In Borexino, diffusion of \po{} into the fiducial volume is a completely negligible process\red{, because on average \po{} can only travel 15 cm via pure diffusion before it decays, while presence of migrated \po{} is found in the center region where \po{} need to travel by around 4.5 m. Therefore the movement of \po{} is mainly driven by convection motion in the liquid scintillator. Because this convection motion is generated by the inhomogeneity and instability of the detector temperature, the migrated \po{}} is time- and space-dependent and hard to model.

These considerations are summarized through the following equation:
\begin{equation}
 R_\text{Po}^\text{tot}(t) = 
 R_\text{Po}^\text{u} \cdot e^{-t/\tau_{Po}} + 
 R_\text{Po}^\text{m}(t)+
 R_\text{Po}^\text{s},
 \label{eq:PoTime}
\end{equation}
where $R_\text{Po}^\text{tot}$ is the observed \po{} rate, \(R_\text{Po}^\text{u}\) is the initial rate of unsupported \po{} decay, \(R_\text{Po}^\text{m}\) is the migrated \po{} decay rate, and \(R_\text{Po}^\text{s}\) is the supported \po{} rate and is equal to the \bi{} rate.

The convective motion of the liquid scintillator was suppressed in a major effort over several years since 2015 \cite{ThermslSimulation2017}.
This included the installation of a passive thermal insulation system around the water tank and an active temperature control system to mitigate seasonal modulations. Our observations showed that 
 the migrated \po{} was significantly reduced in the Borexino Phase-III which starts from 2016 June. 

\red{In Phase-III,} the rate of unsupported \po{} \red{has decayed and is less than 0.1 \cpd{}}. \red{It is also} possible to get a more precise estimate of the rate of supported \po{} by identifying the region inside the fiducial volume, that is least affected by convective motions and thus with minimal migrated \po{}.

The residual migrated \po{} rate in this region can then be treated as a source of systematic uncertainty.
Neglecting the migrated \po{} rate, the statistical uncertainty of the supported term depends on the magnitude of the initial out-of-equilibrium
contamination and on the exposure.
As an example, with 
an unsupported rate of 50 \cpd{} and a supported rate 
of \red{10} \cpd{}, an precision of the order of $10\%$ on the 
supported rate can be achieved in about $8$ months 
of data-taking using the same fiducial volume of 71.3 ton as adopted in \cite{ppchainNature}.

Furthermore, in order to extrapolate the value of \bi{} obtained in the region with minimum \po{} migration rate to the entire fiducial volume, two hypotheses must be verified: the \bi{} spatial distribution must be uniform within the entire fiducial volume and its rate must follow the slow decay rate of \pb{}, since there are no sources of either \bi{} or \pb{} in the liquid scintillator. The residual non-uniformity and instability of the \bi{} rate can also be treated as systematic uncertainties of the \bi{} constraint. 

\subsection{Impact of further scintillator purification}
\label{sec:purification}

The purification campaign of the Borexino scintillator performed in the year 2011 \cite{Purification2012}, with 6 cycles of closed-loop water extraction, significantly reduced the concentrations of several radioactive
contaminants of the scintillator. The \bi{} rate was reduced by a factor $\sim$ 2.3 \cite{ppchainNature,BxNuSol}.
In principle, another cycle of purification could decrease the amount of \bi{} even further without introducing more \po{} in the liquid scintillator. However, in order to keep the amount of migrated \po{} small, thus for the sensitivity to CNO neutrinos to benefit from further purification, the thermal stability of the detector must be maintained.

\begin{figure}[htb]
 \centering
 \includegraphics[width=\columnwidth]
 {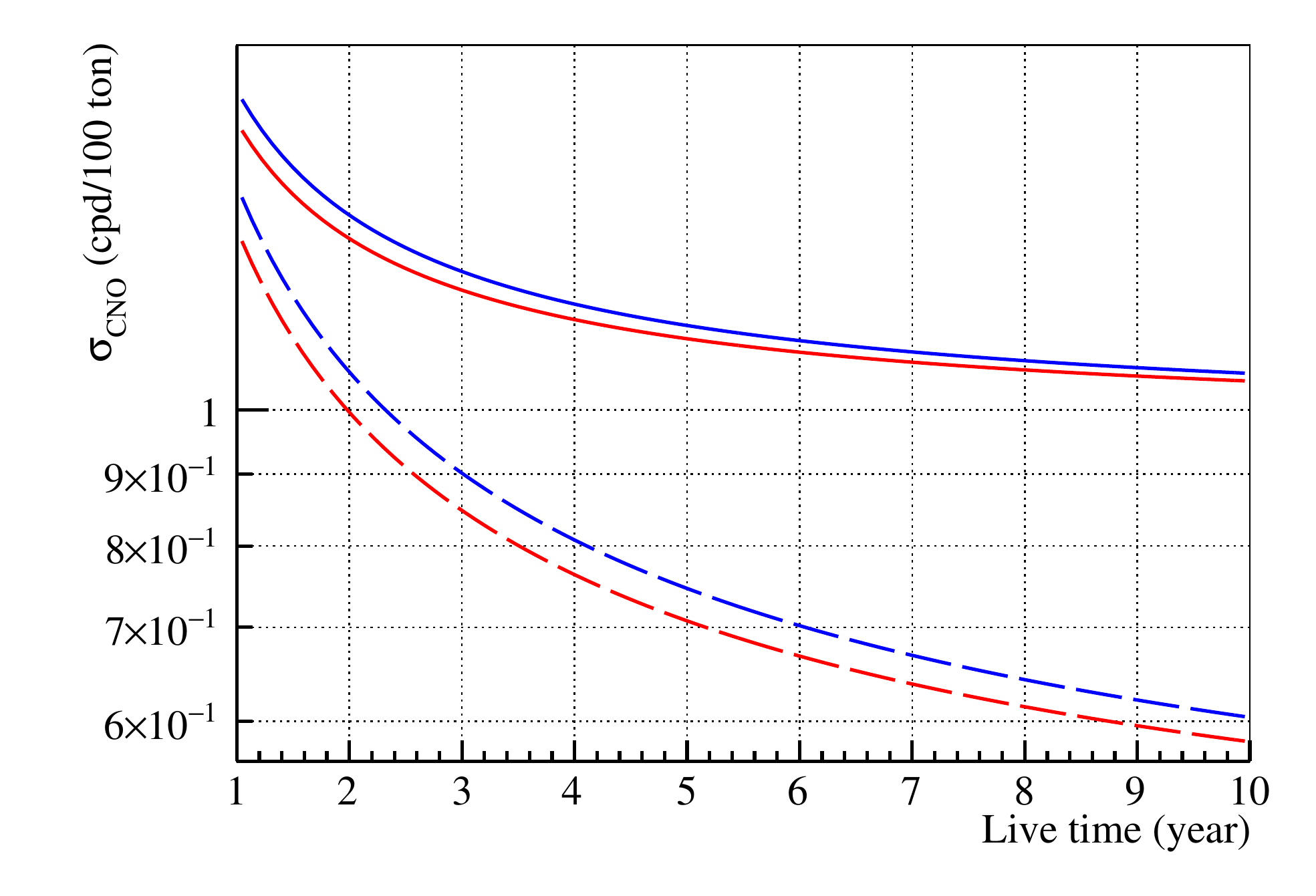}
 \caption[Impact of purification]
  { Expected uncertainty on the CNO interaction rate versus live time assuming a \bi{} constraint uncertainty of 1.5 (dashed line) and 0.5 (solid line) \cpd{} with different values of the \bi{} rate.
  Blue line: \red{10 \cpd{}}. Red line: 5 \cpd{}. From the figure we can see that when \bi{} constraint is strong, the
  purification of liquid scintillator reduces the time needed to reach a given uncertainty on the interaction rate of CNO neutrinos.
 }
 \label{fig:Purification}
\end{figure}

In order to quantify the effect of a reduced rate of \bi{}, we rewrite Eq.~\ref{eq:SigmaCnt} as
\begin{equation}
\sigma_\text{CNO}(R_{\rm Bi}, \mathcal{E}) = \frac{1}{\varepsilon_\text{CNO}} \left(
 \sqrt{\dfrac{\varepsilon_\text{Bi}R_\text{Bi} + a_1}{\mathcal{E}}}
 \oplus a_2 \right),
 \label{eq:bi_expo}
\end{equation}
where \(a_1=1.07\) \cpd{} is the total rate of components other than \bi{} in the ROI,
\(a_2=0.06\) \cpd{} is the total uncertainty of the last three terms in Eq.~\ref{eq:SigmaCnt} and is dominated by \(\tilde{\sigma}_\text{Bi}\), and \(\mathcal{E}\)
is the exposure in day$\times$100 ton. From Eq. \ref{eq:bi_expo} we see that the importance of $\mathcal{E}$ and $R_{\rm Bi}$ is greater when $a_2$ is smaller
than the first term. Thus, in these conditions, reducing the rate of \bi{} is equivalent to increasing the exposure. 
This can also be concluded from Fig.~\ref{fig:Purification}.

%=========================================================
\section{CNO neutrinos in the Standard Solar Model}
\label{sec:SSMimpact}

Despite the small 
contribution of the CNO cycle to the Sun's luminosity, 
a measurement of the CNO neutrino flux would be extremely 
valuable to expand our knowledge of the Sun. 

Experimental measurements of solar neutrino fluxes are
a fundamental test of the SSM, the benchmark of every 
stellar evolution model. 
Possible disagreements between predicted 
and measured neutrino fluxes may highlight issues in the model 
assumptions. Additionally, measurements of solar 
neutrino fluxes can be used to infer 
some properties of the Sun that are used as inputs for
the SSM.

This is particularly interesting in the context 
of the solar metallicity problem.
As mentioned in the introduction, 
SSMs implementing the most recent determinations of the solar 
surface composition (\textit{e.g.}, the low metallicity (LZ) admixture 
described in \cite{AGSS09}),
fail to reproduce helioseismic observations, which agree better with models that assume a higher metal content, as
prescribed by the older determinations 
from \cite{GS98}.
So far, all attempts to understand the origin of this 
discrepancy remain inconclusive. New experimental results on solar neutrinos
could provide key pieces of
information to solve this problem. This applies in particular to the flux of CNO neutrinos, whose value differs by almost 30\% between the predictions of the SSM HZ and LZ models (see Table \ref{tab:exp_rate}).

However, the dependence of predicted solar neutrino fluxes on 
the metallicity and other SSM parameters is non-trivial.
There is indeed a strong degeneracy between the impact of 
metallicity and radiative opacity through their effect on the 
core temperature of the Sun $T_c$ ($T_c$ is determined by the opacity profile,
which in turn 
depends on the chemical composition of the solar interior).
Ultimately, $T_c$ acts as a proxy for the influence of those parameters (\textit{i.e.}, they impact neutrino fluxes via their effect on $T_c$).
This degeneracy makes it difficult to disentangle metallicity 
from opacity with solar neutrino measurements, 
in particular after recent laboratory measurements \cite{Bailey:2015} 
and theoretical advances \cite{Krief2018,Colgan2016,Iglesias2015,Krief2016a,Blancard2012} suggested that the uncertainty of the radiative opacity
might be severely underestimated.
Deviations of the radiative opacity from its nominal values can 
mimic the effect of a higher metal content in the Sun's
core. As it was discussed in, for example, 
\cite{Bahcall:2004yr, OpHoc09, Villante:2010vt},
the agreement of SSM calculations implementing the recent LZ
composition and helioseismic data could be restored by considering 
suitable modifications to the radiative opacity. 
Similarly, neutrino fluxes expected by the HZ SSM can be
obtained by assuming the LZ composition and a larger opacity of 
the solar plasma. 

However, the case of the CNO neutrinos is somewhat different.
In fact, the dependence of the CNO neutrino flux on metallicity
is twofold: as for the other $pp$-chain neutrinos
the CNO flux is indirectly linked to metallicity via
$T_c$, but it also depends directly
on the amount of carbon and nitrogen in the 
core of the Sun.
Considering the metallicity problem, other tests of the SSM
that consider measurements of \ce{^{8}B} and \ce{^{7}Be} 
neutrino fluxes
include a larger amount of information but are greatly affected by uncertainties on
the radiative opacity.

Section~\ref{sec:SSMimpact:CNAbundance} shows that by exploiting this fact, a measurement of the CNO neutrino flux with Borexino can be used to infer the content of carbon and nitrogen in the Sun's core almost independently of the effect of radiative opacity, thus directly probing the solar metallicity.
Sec.~\ref{sec:SSMimpact:HpTest} discusses the impact of a future measurement of CNO neutrinos on the 
discrimination power of a hypothesis test between the HZ and LZ
SSM.

\subsection{CNO neutrinos as CN abundance messengers}
\label{sec:SSMimpact:CNAbundance}

As explained in \cite{Haxton:2008}, the temperature dependence for both the \ce{^{8}B} and CNO neutrino fluxes can be described by a power-law $\Phi_i \propto T_c^{\gamma_i} (i \in \{^{8}\text{B}; \text{CNO}\})$. Therefore, one can build a weighted ratio $\displaystyle \Phi_{\text{CNO}}/(\Phi_{
^{8}\text{B}})^k$, with $\displaystyle k = \gamma_{\text{CNO}}/\gamma_{^{8}\text{B}}$, that is very nearly (considering different uncertainties) independent of $T_c$. This breaks the opacity-composition degeneracy, making it possible to 
infer the abundance of carbon and nitrogen in the Sun in a way
that is almost independent from variations in the SSM parameters
that affect the temperature profile. 

To elaborate, we followed the approach 
of \cite{Haxton:2008, Serenelli:2013}, which adopts the 
practice of factorizing the dependence of the neutrino flux $\Phi_i$
on the model's input parameters $\{\beta_j\}$:
\begin{equation}
  \frac{\Phi_i}{\Phi_i^\text{SSM}} = 
  \prod_{j}^\text{sol} x_j^{\alpha( i, j )}\cdot 
  \prod_{j}^\text{met} x_j^{\alpha( i, j )}\cdot 
  \prod_{j}^\text{nucl}x_j^{\alpha( i, j )},
  \label{eq:prod}
\end{equation}
where the $x_j$ terms are the 
model inputs normalised to their respective nominal SSM value 
(\textit{i.e.}, \ $\beta_j/\beta_j^\text{SSM}$). 
These parameters are typically grouped in three categories: 
(\textit{i}) the \textit{solar} parameters, 
related to the Sun's astrophysical
(age $A_{\odot}$, luminosity $L_{\odot}$)
and non-nuclear physical properties 
(diffusion $D$, radiative opacity $\kappa$),
(\textit{ii}) the \textit{metallicity} parameters (\textit{i.e.},
the abundances of \ce{C}, \ce{N}, \ce{O}, \ce{F}, \ce{Ne}, \ce{Mg}, 
\ce{Si}, \ce{S}, \ce{Ar}, and \ce{Fe}),
and (\textit{iii}) the \textit{nuclear} cross sections of the relevant 
processes, described by the astrophysical $S$-factors. 
The $\alpha( i,j )$ coefficients in the exponents
are normalised logarithmic partial derivatives
of the fluxes with respect to the input parameters $\beta_j$
\begin{equation}
  \alpha(i,j) = \frac{\text{d}\ln(\Phi_i/\Phi_i^\text{SSM})}
  {\text{d}\ln(\beta_j/\beta_j^\text{SSM})},
  \label{eq:logder}
\end{equation}
which are typically provided along with SSM predictions. 

Using Eq.~\ref{eq:prod} to express solar neutrino fluxes, 
the aforementioned weighted ratio between 
 the \ce{^{15}O} and \ce{^{8}B} fluxes reads:
\begin{equation}
  \left(\frac{\Phi_{\ce{^{15}O}}}{\Phi^\text{SSM}_{\ce{^{15}O}}}\right)/
  \left(\frac{\Phi_{\ce{^{ 8}B}}}{\Phi^\text{SSM}_{\ce{^{ 8}B}}}\right)^{k} = 
  \prod_j x_j^{\alpha(\ce{^{15}O}, j) - k\alpha(\ce{^{8}B}, j)}.
  \label{eq:exampleO15}
\end{equation}

To reduce the dependence on $T_c$, the weight $k$ is chosen
such that it minimizes the uncertainty of the quantity
defined in Eq.~\ref{eq:exampleO15}
due to the so-called \emph{environmental} parameters (\textit{i.e.}, those SSM inputs which affect the 
Sun's temperature profile the most \cite{Haxton:2008, Serenelli:2013}). 
These parameters include the \textit{solar}
and \textit{metallicity} parameters discussed above, but without the 
C and N abundances.

When applying this strategy to Borexino we refer, for simplicity, 
to the counting analysis described in Sec.~\ref{sec:counting}.
In this narrow energy window, neglecting insignificant \ce{^{17}F} neutrinos, CNO neutrino events originate from contributions of \ce{^{15}O} and \ce{^{13}N} neutrinos.
We thus define the flux of CNO neutrinos measured by Borexino 
$\Phi_{\ce{CNO}}^\text{BX}$ as a combination of 
$\Phi_{\ce{^{15}O}}$ and $\Phi_{\ce{^{13}N}}$. This can be written as:
\begin{equation}
  \frac{\Phi_{\ce{CNO}}^\text{BX}}{\Phi_{\ce{CNO}}^\text{SSM}} = 
  \xi\, 
  \frac{\Phi_{\ce{^{15}O}}}{\Phi_{\ce{^{15}O}}^\text{SSM}}
   +(1-\xi)\,
  \frac{\Phi_{\ce{^{13}N}}}{\Phi_{\ce{^{13}N}}^\text{SSM}},
  \label{eq:flux_xi}
\end{equation}
where $\xi$ is the ratio between the event rate of 
\ce{^{15}O} neutrinos and all CNO neutrinos in
the ROI, as expected from the SSM:
\begin{align}
 \xi \equiv r_{\ce{^{15}{O}}}^\text{SSM}/ 
  (r_{\ce{^{15}{O}}}^\text{SSM}+r_{\ce{^{13}{N}}}^\text{SSM}) = 
  r_{\ce{^{15}O}}^\text{SSM}/r^\text{SSM}_{\ce{CNO}} = 0.764.
\end{align}
Using the logarithmic derivatives of \ce{^{15}O} and
\ce{^{13}N} neutrinos from the B16-SSM \cite{B16SSM},
we can describe the dependence of the measured CNO neutrino flux
on the inputs of the SSM as
\begin{equation}
  \alpha(\ce{CNO}^\text{BX}, j) =
       \xi    \,\alpha(\ce{^{15}O}, j) + 
       (1-\xi)\,\alpha(\ce{^{13}N}, j).
  \label{eq:alphacn}
\end{equation}
Having computed those derivatives, we can express Eq.~\ref{eq:exampleO15} using a CNO measurement by Borexino and the \ce{^{8}B} neutrino flux measured by Super-Kamiokande \cite{Abe2016}. In the case of Borexino, the value of $k$, obtained from a minimization of the uncertainty from the \textit{environmental} parameters, is found to be $0.716$. To explicitly show the dependence on all SSM inputs: 
\begin{multline}
\frac{\Phi_{\ce{CNO}}^\text{BX}}
{\Phi_{\ce{CNO}}^\text{SSM}}
/
\left[\frac{\Phi_{\ce{^{8}B}}}{\Phi_{\ce{^{8}B}}^\text{SSM}}\right]
^{0.716} 
 = \\ 
 x^{0.814}_{\ce{C}}\,x^{0.191}_{\ce{N}}\,x_{D}^{0.184} \times \\
 \times 
 \left[ 
   x_{L_\odot   }^{ 0.618}\, x_{\kappa_{a}}^{0.023}\,
   x_{\kappa_{b}}^{-0.048}\, x_{A_\odot   }^{0.274}
 \right] \times \\
%----------------------
 \times 
 \left[ 
   x^{ 0.005}_{\ce{O} }\, x^{-0.004}_{\ce{Ne}}\,
   x^{-0.003}_{\ce{Mg}}\, x^{ 0.001}_{\ce{Si}}\,
   x^{ 0.001}_{\ce{S} }\, x^{ 0.001}_{\ce{Ar}}\, 
   x^{ 0.004}_{\ce{Fe}} 
 \right]\times \\
%--------------------
 \times 
 \left[
   x_{S_{11}}^{-0.820}\, x_{S_{33}}^{ 0.324}\, x_{S_{34}}^{-0.647}\,
   x_{S_{e7}}^{ 0.715}\, x_{S_{17}}^{-0.736}\, x_{S_{114}}^{0.978} 
 \right],
 \label{eq:CNB8}
\end{multline}
where the contributions of the environmental parameters\footnote
{
  The radiative opacity is represented
  by two parameters, namely $\kappa_{a}\equiv 1+a$ and 
  $\kappa_b \equiv 1+b$, which describe
  the variations of the solar opacity profile
  in terms of the parameters $a$ and $b$ defined in 
  \cite{B16SSM}.
}
are grouped in the second and third rows on the right-hand side, 
while the contributions of nuclear cross sections are 
in the fourth row.
Following \cite{Haxton:2008, Serenelli:2013}, 
the exponents in Eq.~\ref{eq:CNB8} 
were obtained using the values of the 
logarithmic derivatives of the solar neutrino fluxes with
respect to the SSM inputs used by the HZ SSM \cite{B16SSM}.

Assuming, for simplicity, that the \ce{C} and \ce{N} abundances 
are modified by the same factor (\textit{i.e.}, 
$x_{\ce{C}} = x_{\ce{N}} \equiv 
(N_{\ce{C}}+N_{\ce{N}})/(N_{\ce{C}}^\text{SSM} + 
N_{\ce{N}}^\text{SSM})$, where $N_{\ce{C}}$ and $N_{\ce{N}}$ indicate
the number density of C and N with respect to hydrogen)
and noticing in Eq.~\ref{eq:CNB8} that the sum of the exponents of
$x_{\ce{C}}$ and $x_{\ce{N}}$ is $0.814+0.191\simeq 1$, we
can invert the formula, obtaining:
\red{\begin{multline}
 \frac{N_{\ce{C}}+N_{\ce{N}}}
 {N_{\ce{C}}^\text{SSM}+N_{\ce{N}}^\text{SSM}}= 
 \left(\frac{\Phi_{\ce{^{8}B}}}
 {\Phi_{\ce{^{8}B}}^\text{SSM}}\right)^{-0.716}
 \times \\
 \times \frac{R_{\ce{CNO}}^\text{BX}}{R_{\ce{CNO}}^\text{SSM}}
 \times
 \left[
 1 \pm 0.5\% (\text{env}) \right. \\
 \left. \pm\, 9.1\% (\text{nucl}) \pm 2.8\% (\text{diff})
\right],
\label{eq:CN}
\end{multline}}
where $R_{\ce{CNO}}^\text{BX}$ is the rate measured by Borexino and proportional to the flux $\Phi_{\ce{CNO}}^\text{BX}$.
The quoted uncertainties were obtained by propagating the uncertainties 
of the environmental, nuclear, and diffusion parameters used 
in the SSM. The current uncertainty
on the measurement of the boron neutrino flux is about 2\% \cite{Bergstrom:2016cbh}. \red{The dominant contributions to the uncertainty budget besides the CNO neutrino rate are from nuclear reactions,
with the largest coming from $S_{114}$ ($7.3\%$), $S_{34}$
($3.4\%$), and $S_{17}$ ($3.5\%$).}

The low level of the environmental contribution to the total 
error budget (which takes into account all SSM parameters except nuclear 
reactions and diffusion coefficients) demonstrates 
that a future \ce{CN} flux measurement will be converted into an
almost direct determination of the C+N content of the solar core, 
as discussed in \cite{Haxton:2008, Serenelli:2013}.

\begin{figure}[h]
 \centering
 \includegraphics[width=\columnwidth]{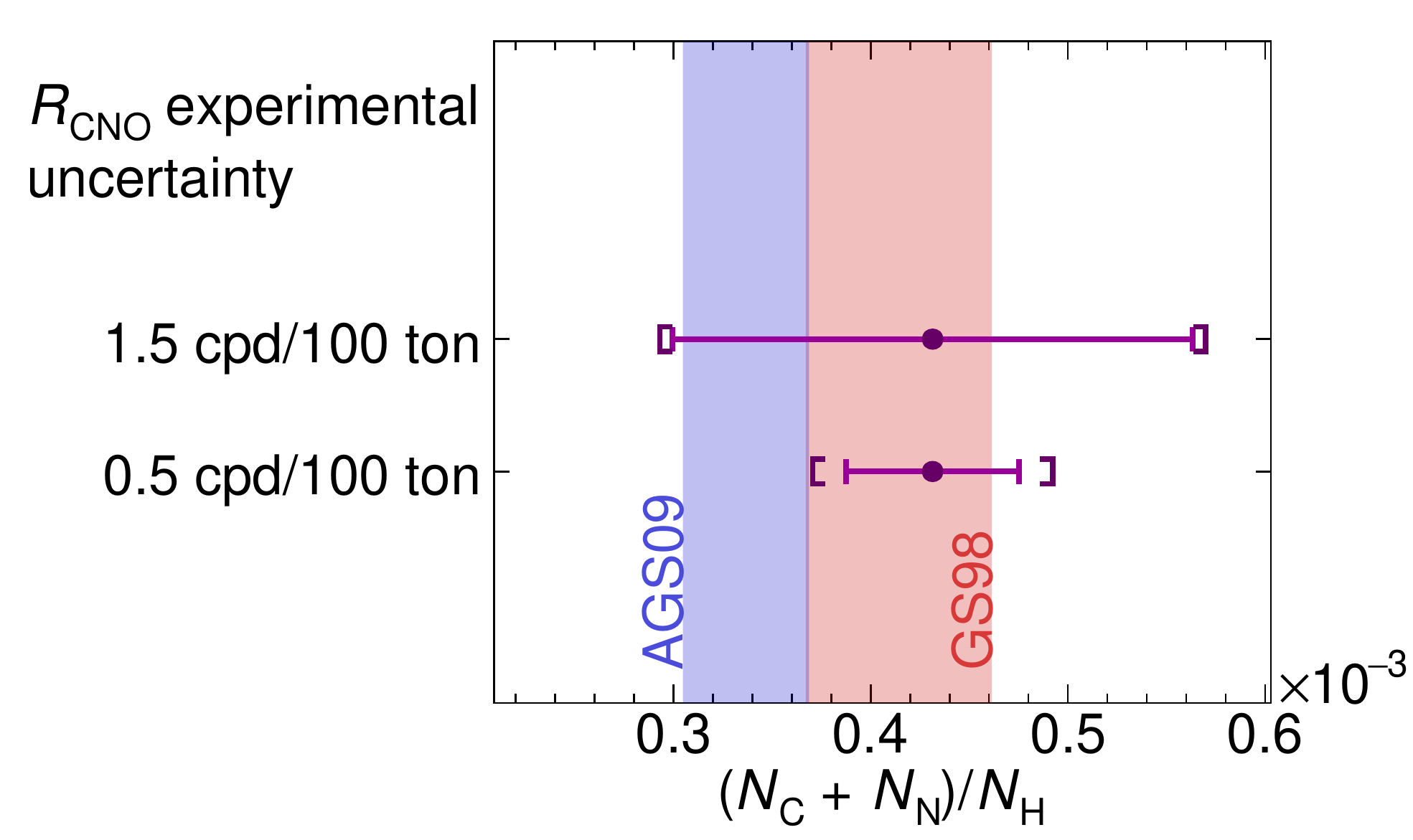}
 \caption{
 Projected uncertainty in the determination of the (C+N) abundance from a measurement of CNO neutrinos under different precision scenarios. 
The error bars indicate the error budget due to the experimental precision in the measurement of CNO solar neutrinos, while the overall uncertainty---accounting for the limited precision of the SSM inputs and for the \ce{^{8}B} solar neutrino uncertainty---is enclosed in square brackets. 
These results have been obtained assuming the rate expected from the HZ SSM and using the result of \cite{Bergstrom:2016cbh} for the \ce{^{8}B} flux. The red and blue bands show the 68\% confidence interval of the abundance of carbon and nitrogen reported in the GS98 \cite{GS98} and in the AGS09 \cite{AGSS09} catalogues respectively. }
 \label{fig:cnAbundance}
\end{figure}

The impact of the experimental and model uncertainties in 
the determination of the C+N abundance inferred from a CNO neutrino 
measurement, assuming the rate predicted by the HZ SSM, is represented in Fig.~\ref{fig:cnAbundance}. 
While a \SI{1.5}{} \cpd{} precision is translated, according to 
Eq.~\ref{eq:CN},
to a precision that is about three to four times larger than the precision 
of the GS98 and AGSS09 catalogues, a future measurement of
$R_{\ce{CNO}}$ with a precision
$\simeq$ \SI{0.5}{} \cpd{} (achievable by next generation experiments
\cite{Franco:2016, THEIA:2019,Beacom2017}) will be able to constrain the C+N fraction of the 
Sun with an uncertainty $\simeq 15\%$, which is comparable to 
the precision of the current estimations obtained from measurements
of the photosphere. 

\subsection{High versus Low Metallicity Standard Solar Model}
\label{sec:SSMimpact:HpTest}
If one considers the current treatment of the opacity 
and of its corresponding uncertainties in the SSM as reliable, 
then the measurements of neutrino fluxes due to the \pp{} chain can 
also be used to test models of the composition of the Sun.

In the results of Borexino Phase-II \cite{ppchainNature}, the measurement of the fluxes of \ce{^{7}Be} and \ce{^{8}B}
neutrinos were used to this end because of their strong dependence 
on the Sun's core temperature, which is affected by the metal content as described above. 
Including CNO neutrinos in a similar test is particularly 
interesting since their flux depends on the carbon and nitrogen abundances in the Sun.
This strong dependence on the SSM parameters leads to 
a $30\%$ difference between the predictions of 
the CNO neutrino flux of the HZ and LZ SSMs (See Table~\ref{tab:exp_rate}).

In order to assess the relevance of a future CNO flux determination in this context,
we \red{performed a frequentist hypothesis test 
analogue to the one described in \cite{ppchainNature} and considered the HZ and LZ SSMs as the alternative hypothesis (\(H_1\)) and the null hypothesis (\(H_0\)), respectively.}
Namely, we assumed a given model and
determined the $p$-value against the alternative hypothesis
when the ${^{8}{\rm B}}$, ${^{7}{\rm Be}}$, and CNO neutrino
rates were measured with prescribed accuracies.
For ${^{8}{\rm B}}$ and ${^{7}{\rm Be}}$ neutrinos, 
we adopted the rate uncertainties from \cite{ppchainNature}. This makes it possible to directly compare our results with the one from \cite{ppchainNature} that did not incorporate CNO neutrinos in the analysis.
To simulate future CNO neutrino measurements, we used uncertainties of $1.5$ and $0.5$ \cpd{}, as discussed in the previous sections. 
 
\begin{figure}[htb]
 \includegraphics[width=0.50\textwidth] 
 {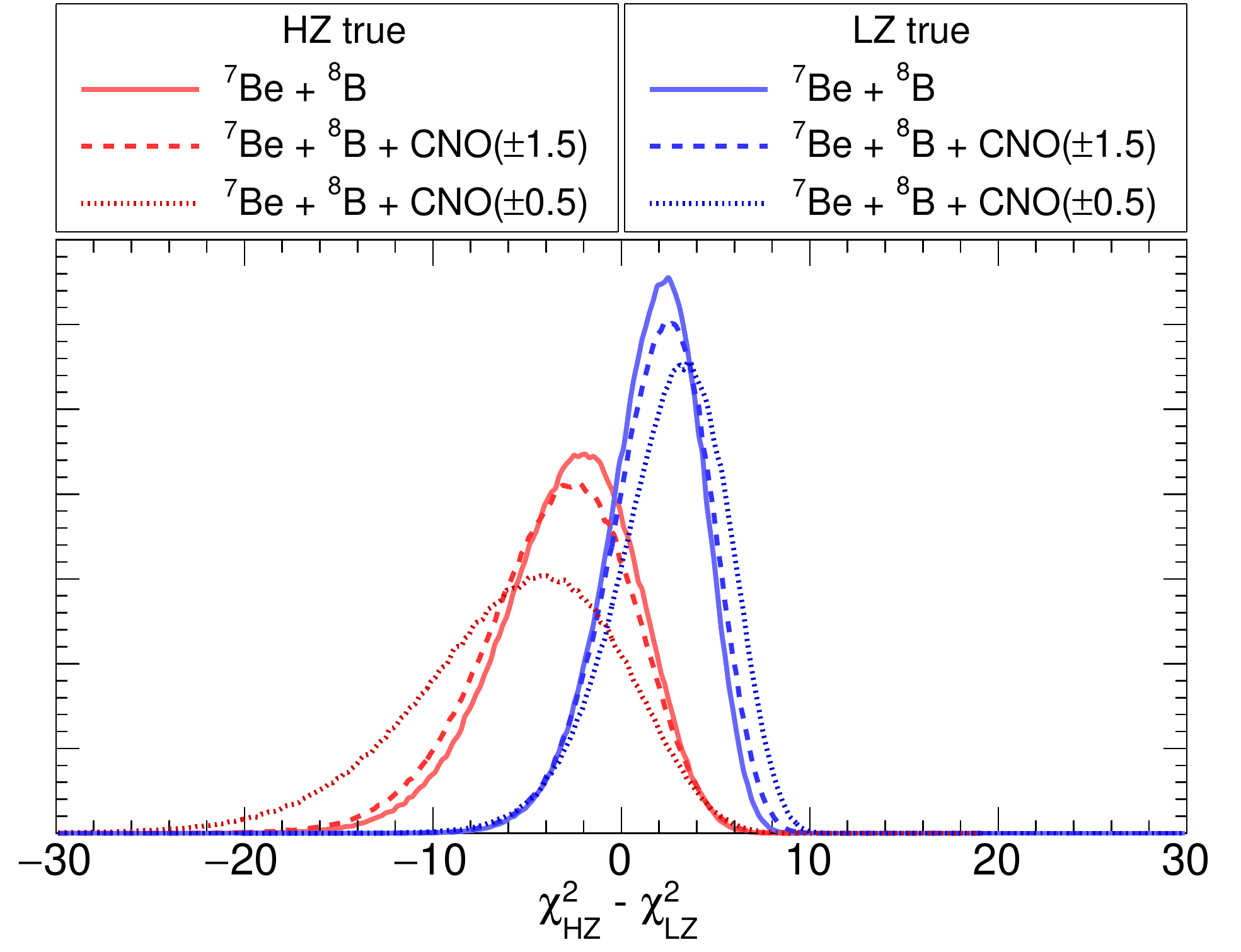}
 \caption{
 PDFs of a test statistics suited
 to perform a hypothesis test between the HZ and LZ SSM. 
 The solid-line PDFs indicates the expected distributions
 obtained considering only \ce{^{8}B} and \ce{^{7}Be}
 neutrinos obtained with the same precision as reported by Borexino in \cite{ppchainNature}, while the distributions with dashed 
 and dotted lines are obtained including in the analysis 
 a future measurement of CNO neutrinos with an precision of 
 \SI{1.5}{} and \SI{0.5}{} \cpd{}, respectively.
 }
 \label{fig:tstat_pdf}
\end{figure}

Our results are shown in Fig.~\ref{fig:tstat_pdf}, 
where we plot the distributions of the test statistic defined as the 
difference between the $\chi^2$ computed for HZ and LZ predictions 
including both model and experimental uncertainties.
The more the distributions are separated, 
the higher the probability of discriminating among the different
hypotheses is. 

When taking into account values for the \ce{^{8}B} and \ce{^{7}Be}
neutrino fluxes as predicted by the HZ SSM and measured with the same precision as in \cite{ppchainNature}, 
the median $p$-value for the LZ predictions is $0.057$, corresponding 
to a $1.6\sigma$ significance in the exclusion of the wrong 
hypothesis.
When including a measurement of CNO neutrinos with an uncertainty
of \SI{1.5}{} \cpd{}, the median discriminatory power for the 
LZ SSM hypothesis does 
not change significantly ($p\text{-value} = 0.047$, $1.7\sigma$ 
significance), as the experimental uncertainty is $2$--$3$ times 
larger than the model precision. 
A larger increase in significance can be obtained
if a \SI{0.5}{} \cpd{} precision in the determination of the 
CNO neutrino flux is achieved. 
In this case, assuming the HZ SSM predictions, 
the median $p$-value for the LZ SSM is $0.016$ ($2.1\sigma$).
Similar significance levels were found when using the values of $\Phi_{\ce{^{8}B}}$ and $\Phi_{\ce{^{7}Be}}$ measured by Borexino in \cite{ppchainNature}.

This test relies on stronger assumptions as the determination 
of C+N abundance presented in the previous section. It also considers a 
simplified ``binary'' hypothesis system where either the HZ and LZ
SSM is assumed to be correct, with no possible alternative explanation.
However, its discriminatory power is better than the one achievable 
from the inferred C+N abundance for the simple reason that while in 
the latter the impact of metallicity on the core temperature is 
cancelled out, in this study it is the dominant source of information,
driving the sensitivity of the test through $\Phi_{\ce{^{8}B}}$ and 
$\Phi_{\ce{^{7}Be}}$.
In this context, a measurement of CNO neutrinos 
is expected to have a sensible impact on the test only if the 
experimental precision approaches model uncertainty (\textit{i.e.}, $\sigma_{\rm CNO}\simeq$\SI{0.5}{} \cpd{}).
The precision of the solar model predictions indeed poses a 
strong limit to the overall sensitivity of such a test.
Currently, the largest sources of uncertainty---besides the plasma opacity and the carbon abundance---are the $S_{17}$ (\ce{^{8}B}), 
$S_{34}$ (\ce{^{7}Be}), and $S_{114}$ (\ce{CNO}) astrophysical 
$S$-factors \cite{B16SSM}.
Therefore, a reduction of the nuclear cross section uncertainties is crucial 
to allow for a more significant test of the SSM based on solar neutrino 
measurements.

%=============================================

\section{Conclusions }
\label{sec:CNOStrat:Concl}

Borexino has a a strong potential to detect CNO neutrinos due to its unique radiopurity, intrinsically low cosmogenic \ce{^{11}C} rate and successful TFC method for its further reduction, reduction of external gammas by passive detector shielding, as well as comprehensive detector and background modeling.

A detailed sensitivity study was performed which 
showed that if the backgrounds are constrained, the 
bulk of the sensitivity to the CNO neutrino signal 
can be established with a simple counting analysis. 
The multivariate analysis only mildly improves
this sensitivity when the uncertainty on the 
background rates is relatively loose. 
The statistical uncertainty on the CNO rate depends on
the uncertainty in the determination of the \pep{} neutrinos and \bi{}
backgrounds. \red{If \(R_{\rm Bi}\) and \(R_{\pep{}}\) can be constrained
with a 1.5 \cpd{} and 0.04 \cpd{} 
precision, respectively, the expected uncertainty of CNO neutrino rate is 1.2 \cpd{}, and the expected median discovery significance to CNO neutrinos is $4.1\sigma$ (\(3.0\sigma\)) for the CNO flux assuming 
the HZ (LZ) SSM.}

The sensitivity does not improve significantly
if a new purification campaign were to further reduce the \bi{}
content, unless the absolute precision of the \bi{} constraint is improved. This improvement is not guaranteed by the reduction of the \bi{} content alone and is mainly related to the capability of stabilizing the detector temperature and suppressing convective motions. However, for a given uncertainty on the \bi{} constraint and if the uncertainty on the rate of CNO neutrinos is dominated by statistical fluctuations, a reduction of \bi{} would shorten the time needed to achieve a given precision on the CNO neutrino interaction rate.

In order to break the correlations, the \pep{} and \bi{}
background rates must be independently constrained.
The rate of \pep{} neutrinos can be constrained by exploiting its
relation with the \pp{} neutrino rate as already done in the analysis presented in \cite{ppchainNature}.
The uncertainty on the \pep{} rate can be 
 reduced to the level of around one percent by imposing an additional 
constraint based on solar luminosity measurements. 

The \bi{} isotope is supported by their long-lived
parent \pb{} ($\tau_\text{Pb}=32.2$ years).
The most promising method to constrain the \bi{}
background relies on a measurement of the daughter isotope \po{}, 
which is expected to be in equilibrium with the \pb{} when the liquid scintillator is thermally stable. 
\po{} events can efficiently be tagged using pulse-shape
discrimination. Since mid-2015, the Borexino collaboration has been focusing its effort on the thermally stabilizing the detector.

By combing the measurements of \(\Phi_{\ce{^8B}}\) and \(\Phi_{\rm CNO}\), the carbon and nitrogen contents of the solar core can be determined independently of the opacity of solar plasma. Assuming $R_{\ce{CNO}}$ is measured with an uncertainty
$\simeq$ \SI{0.5}{} \cpd{}, the C+N fraction of the 
Sun can be determined within $\simeq 15\%$, which is comparable to 
the precision of the current estimations obtained from measurements
of the photosphere, and limited by the nuclear cross section precision. 

The neutrino fluxes depend on the chemical composition of the Sun, and can be used to discriminate between SSM HZ and LZ models.
Combining the measurements of \(\Phi_{\be{}},\,\Phi_{\ce{^8B}}\), and \(\Phi_{\rm CNO}\), assuming the CNO rate to be determined at a precision of 0.5 \cpd{}, the median discrimination sensitivity is 2.1\(\sigma\) and limited by the uncertainties of SSM.

%========================================================

\acknowledgement
The Borexino program is made possible by funding from Istituto Nazionale di Fisica Nucleare (INFN) (Italy), National Science Foundation (NSF) (USA), Deutsche Forschungsgemeinschaft (DFG) and\\
Helmholtz-Gemeinschaft (HGF) (Germany), 
Russian Foundation for Basic Research (RFBR) (Grants No. 16-29-13014ofi-m, No. 17-02-00305A, and No. 19-02-00097A) 
and Russian Science Foundation (RSF) (Grant No. 17-12-01009) (Russia), 
and Narodowe Centrum Nauki (NCN) (Grant No. UMO 2017/26/M/ST2/00915) (Poland).

We gratefully acknowledge the computing services of \\
Bologna INFN-CNAF data centre and U-Lite Computing Center and Network Service at LNGS (Italy), 
and the computing time granted through JARA on the supercomputer JURECA \cite{jureca} at Forschungszentrum J\"ulich (Germany). 
This research was supported in part by PLGrid Infrastructure. 
We acknowledge also the generous hospitality and support of the Laboratori Nazionali del Gran Sasso (Italy).

\bibliography{cno}
\bibliographystyle{bx-epj-bibstyle}

\end{document}